\documentclass[12pt]{article}
\usepackage{amssymb}
\usepackage{amsfonts}
\usepackage{amsmath}
\usepackage[mathscr]{eucal}
\usepackage{amssymb}
\usepackage{amsthm}
\usepackage{bm}
\usepackage{graphicx}
\usepackage[english]{babel}
\usepackage[T2A]{fontenc}
\usepackage[utf8]{inputenc}
\usepackage{cite}

\def\theequation{\arabic{section}.\arabic{equation}}
\newcommand{\be}{\begin{equation}}
\newcommand{\ee}{\end{equation}}
\newcommand{\bea}{\begin{eqnarray}}
\newcommand{\eea}{\end{eqnarray}}

\newcommand{\lb}{\label}
\newcommand{\p}[1]{(\ref{#1})}

\newcounter{rown}

\begin{document}
\begin{titlepage}

\vspace*{0.1cm}

\begin{center}
{\LARGE\bf Continuous spin superparticle }

\vspace{0.3cm}

{\LARGE\bf in $4D$, ${\cal N}=1$ curved superspace}

\vspace{1.5cm}

{\large\bf I.L.\,Buchbinder$^{1,2,3}$\!\!,\ \ \
S.A.\,Fedoruk$^1$}

\vspace{1.5cm}

\ $^1${\it Bogoliubov Laboratory of Theoretical Physics,
Joint Institute for Nuclear Research, \\
141980 Dubna, Moscow Region, Russia}, \\
{\tt buchbinder@theor.jinr.ru, fedoruk@theor.jinr.ru}

\vskip 0.5cm

\ $^2${\it Center of Theoretical Physics,
Tomsk State Pedagogical University, \\
634041, Tomsk, Russia}

\vskip 0.5cm

\ $^3${\it National Research Tomsk State University,\\
\it 634050, Tomsk, Russia}\\

\end{center}

\vspace{2cm}

\nopagebreak

\begin{abstract}
\noindent We present a new particle model that describes the dynamics
of a $4D,$ $\mathcal{N}{=}\,1$ continuous spin particle in $AdS_4$
superspace and is a generalization of the
continuous-spin superparticle model in flat $4D$, $\mathcal{N}{=}\,1$
superspace proposed in {\tt 2506.19709\,[hep-th]}. The model is
described by $4D$, $\mathcal{N}{=}\,1$ superspace coordinates
together with commuting spinor additional variables, which are
inherent ingredients of continuous spin models. The Lagrangian and
the system of four bosonic and four fermionic phase space
constraints are derived. The consistency condition for constraints
imposes a restriction on supergeometry to be $AdS$ superpace. It is
shown that the bosonic constraints are first-class constraints. A covariant
procedure based on the use of additional variables is developed to
divide the four fermionic constraints into first and second classes.
It is proved that, unlike the flat case, only one fermionic constraint is
a first-class constraint, while the other three are second-class constraints. In
the flat limit, one of these second-class constraints becomes a
first-class one.
\end{abstract}

\vspace{2cm}


\phantom{Keywords: }  mechanics

\newpage

\end{titlepage}

\setcounter{footnote}{0}

\setcounter{equation}{0}

\section{Introduction}

\setcounter{equation}{0} One-dimensional models describing the
dynamics of relativistic particles in background spaces, in addition to
being interesting in themselves due to many remarkable properties, are
a testing ground for constructing more complicated gauge field
theories and for investigating background geometry. These
models have a variety of global and local (super)symmetries and can
serve as a basis for developing new promising multidimensional
(super)gauge theories. At present, the investigation of relativistic
particle models has developed into a fairly vast trend of research
in theoretical and mathematical physics. The number of publications
on this topic apparently amounts to several hundred.

Among the various models of relativistic particles, one can single
out the superparticle models whose world lines lie in flat or curved
superspaces.  Such models are widely discussed in the literature,
and many of their classical and quantum aspects have been studied in
detail. Within the extensive literature on superparticle models, we
will only mention those
publications that are somehow relevant to our
present paper
\cite{Cas-1976,BSch,AzLuk-1982,Sieg-1983,Lusanna-Milewski_1984,Gupta-Bleuler,Gupta-Bleuler-1,
Witten_1986,Shapiro-Taylor_1,Shapiro-Taylor_2,BHT-1987,Sieg-1988,Shapiro-Taylor_review,
BLPS-2000,BILS-2002,BdAIL-2003,FIL-2006,BSa,KKR,KR}. Later, we will
refer to these papers in more detail at  the appropriate points of our
work. Note, however, that all these papers study only
superparticle models with definite superspins or superhelicities.
The bosonic sector of the above models in flat space corresponds to
irreducible finite spin (or helicity) representations of
the Poincar\'e group.

Among a number of papers on superparticles we especially mention the
papers
\cite{Lusanna-Milewski_1984,Witten_1986,Shapiro-Taylor_1,Shapiro-Taylor_2,BdAIL-2003}
(see also the review \cite{Shapiro-Taylor_review}) where
superparticle models were studied in the curved superspace. Let us
also note the papers \cite{BLPS-2000,BILS-2002,KKR,KR}, which
considered the superparticle in the $AdS$ superspace. The models
under consideration have a remarkable gauge structure that in the
canonical approach manifests itself in the appearance of first-class
bosonic constraints and specific fermion constraints with a not
evident and non-trivial division into first and second classes.
Note that for Lorentz-covariant division of such constraints it is necessary to introduce additional
 bosonic variables, which are often absent in the original Lagrangian.
It is worth pointing out that the consistency condition for
first-class constraints imposes a restriction on the geometry of
curved superspace in the form of supergravity equations. It was also
found that a similar result occurs in superstring theory in the
supergravity background (see, e.g., the review
\cite{Shapiro-Taylor_review}).

Until recently, the continuous spin representations of the Poincar\'e group have not
been studied in the context of superparticle models.
Recently in \cite{BF-25} we constructed a novel relativistic
particle model that for the first time describes the propagation of
a continuous spin superparticle in $4D$, ${\cal N}{=}\,1$ flat
superspace. An approach was based on a supersymmetric generalization of
our earlier works where the bosonic continuous spin particle models
were constructed in flat and curved space-times
\cite{BFIR,BFIK-24-1,BFIP}.\footnote{Relatively recently, a
direction has emerged in field theory associated with the field
description of irreducible massless representations of the
Poincar\'{e} group with continuous spin. Among a large number of
works in this direction, we note only some papers of the last few
years (see. e.g.,
\cite{Metsaev-2018,Metsaev-2023,Metsaev-2025,Metsaev-2025_1,BKT,BFIK-24-2,
Schuster-Toro_1,Schuster-Toro_2,Schuster-Toro_3,Schuster-Toro_4,Schuster-Toro_5}
and references therein). The papers \cite{BFIR,BFIK-24-1,BFIP,BF-25}
can be considered as a natural part of this direction.
Supersymmetric continuous spin field models were discussed, e.g., in the
recent papers
\cite{BGK-2019,BKSZ-2019,Metsaev-2019,N-2020,N-2020_1,BFIK}.} A
remarkable feature of continuous spin models is that in such models
additional bosonic variables are an obligatory part of
the coordinates. Using the commuting spinors as these additional
coordinates introduced in works \cite{BFIR,BFIK-24-1,BFIP}
allowed us, in the superparticle model \cite{BF-25}, to
divide in a natural way the fermion constraints in phase space into
first- and second-class constraints. In the present paper, we propose
a generalization of the model \cite{BF-25} and formulate a new
Lagrangian continuous spin superparticle model in the curved
$4D$, $\mathcal{N}{=}1$ superspace corresponding to the
supergravity background (see, e.g., \cite{WessBagger,Ideas}). As we
will see, an additional bosonic spinor coordinates
play an extremely non-trivial role in the division of fermionic
constraints into first and second class and, besides, the model
turns out to be consistent only in AdS superspace.
The aspects of $AdS$ symmetry needed for our purposes were considered in \cite{IvSor},
where $AdS$ superspace was first presented in explicit form
\footnote{For other pioneering studies on $AdS$ supersymmetry see in \cite{Keck,Zumino,IvSor-a,IvSor-b}.}.

The paper is organized as follows. In Section\,2 we briefly describe
the continuous spin model in flat $4D$, ${\cal N}{=}\,1$
superspace derived earlier in \cite{BF-25}. Section\,3 is devoted to
problems of generalizing the results of \cite{BF-25} to $4D$, ${\cal
N}{=}\,1$ curved superspace using the technique of differential
supergeometry \cite{Ideas}. It is pointed out that a direct
generalization turns out to be too naive and leads to different
inconsistences and requires a modification. In Section\,4 we first
show that the theory under consideration can consistently be
formulated only in $AdS$ superspace and second we derive  consistent
bosonic and fermionic constraints and a consistent Lagrangian. It
is shown that the modified bosonic constraints form the first-class
constraints. As to fermionic constraints, their class is not evident
from the very beginning. We have developed a procedure for dividing the
fermionic constraints into first- and second-class constraints using
additional bosonic spinor variables that are inherent
in continuous spin models. It is shown that four fermionic
constraints are divided into three second-class constraints and one
first-class constraint. This is significantly different from the
superparticle model in flat superspace, where four fermion
constraints are divided into two second-class constraints and two
first-class constraints. In the flat limit, one of the fermionic
second-class constraints transforms into a first-class constraint. In
Section\,5, using the first-class fermionic constraint as a
generator of gauge transformations in phase space, we derive
$\kappa$-symmetry transformations for the continuous spin superparticle
model in curved superspace. Note that the auxiliary variables are also
transformed under $\kappa$-symmetry transformations in the case
under consideration. The Summary is devoted to the discussion of the results
obtained. Some auxiliary technical details are given in the
Appendices.

\setcounter{equation}{0}

\section{Continuous spin superparticle in flat superspace}

Before constructing the superparticle model in curved superspace, we
describe the infinite spin particle in $4D$ flat target superspace. To do this,
we use the anticommuting  Weyl spinor fields $\theta^{\alpha}$,
$\bar\theta_{\dot\alpha}$, which, together with the bosonic coordinates
$x^{a}$, form the $4D$, ${\cal N}{=}\,1$ flat superspace with
the coordinates $z^A=(x^a,\theta^{\alpha},\bar\theta_{\dot\alpha})$. In
describing the continuous spin particle we use additional spinor
coordinates $\xi^\alpha$,\, $\bar\xi^{\dot\alpha}=(\xi^\alpha)^*$
and $\pi_\alpha$,\, $\bar\pi_{\dot\alpha}=(\pi_\alpha)^*$. The
latter of them play the role of momenta for the former.

In \cite{BF-25}, we derived the target space
supersymmetric-invariant Lagrangian in the form
\begin{equation}
\label{L-super-1} L_{0} \ = \ \frac{1}{2e}\,W^{\,a} W_{a} \ +  \
\nabla\xi^{\alpha} \pi_{\alpha}  \   +  \ \bar\pi_{\dot\alpha}
\nabla{\bar\xi}^{\dot\alpha} \  - \
(\lambda+\tilde\lambda)\bm{\mu}\,.
\end{equation}
Here the four-vector
\begin{equation}
\label{Omega-1}
W^{\,a} \ := \ \omega^a+ \lambda (\xi\sigma^a \bar\xi ) + \tilde\lambda (\bar\pi\tilde\sigma^a \pi )
\end{equation}
is a continuous spin generalization of the Cartan $\omega$-superforms
\begin{equation}
\label{om}
\omega^a \ = \ \dot x^a  +i\theta\sigma^a\dot{\bar\theta}-i\dot\theta\sigma^a{\bar\theta}\,,
\end{equation}
which are invariant with respect to supersymmetry transformations
\begin{equation}
\delta x^a = i\theta\sigma^a{\bar\varepsilon}-i\varepsilon\sigma^a{\bar\theta}\,,
\qquad  \delta\theta^{\alpha}=\varepsilon^{\alpha}\,,\qquad  \delta\bar\theta_{\dot\alpha}=\bar\varepsilon_{\dot\alpha}
\end{equation}
with the anticommuting global parameters $\varepsilon^{\alpha}$,
$\bar\varepsilon^{\dot\alpha}=(\varepsilon^{\alpha})^*$. Note that
the bosonic auxiliary coordinates and momenta are inert with respect
to supersymmetry transformations: $\delta\xi^{\alpha}=0$,
$\delta\bar\xi^{\dot\alpha}=0$, $\delta\pi_{\alpha}=0$,
$\delta\bar\pi_{\dot\alpha}=0$.

In \p{L-super-1}, $x^a$, $\xi^{\alpha}$, $\bar{\xi}^{\dot{\alpha}}$,
$\pi_\alpha$, $\bar\pi_{\dot\alpha}$ are $1D$ fields, where $\tau$
is the evolution parameter: $x^a=x^a(\tau)$,
$\xi^{\alpha}=\xi^{\alpha}(\tau)$, etc. We use the standard notation
for $\tau$-derivatives: $\dot x^a$, $\dot \xi^{\alpha}$, $\dot
{\bar\xi}^{\dot\alpha}$. The parameter $\bm{\mu}\in\mathbb{R}$ in
\p{L-super-1} is a nonzero constant and  $e(\tau)$, $\lambda(\tau)$,
$\tilde{\lambda}(\tau)$, $A(\tau)$ are the Lagrange multipliers. In
addition, the quantities
\begin{equation}
\label{nabla-xi}
\nabla\xi^{\alpha}=\dot \xi^{\alpha} +iA\xi^{\alpha}\,,\qquad
\nabla{\bar\xi}^{\dot\alpha}=\dot {\bar\xi}^{\dot\alpha}-iA{\bar\xi}^{\dot\alpha}
\end{equation}
are the $\mathrm{U}(1)$-covariant derivatives.

The model with the Lagrangian \p{L-super-1} is described by
the fermionic constraints
\begin{equation}\lb{D-constr}
D_{\,\alpha}=ip_{\alpha}-(\sigma^a\bar\theta)_\alpha p_a\approx 0\,,
\qquad   \bar D_{\,\dot\alpha}=-(D_{\,\alpha})^*=  i\bar p_{\dot\alpha}+(\theta\sigma^a)_{\dot\alpha}p_a\approx 0\,,
\end{equation}
and the bosonic constraints
\begin{eqnarray}
l_0 &=& p_a p^a \ \approx \ 0 \,, \label{const-sp}\\ [5pt]
l &=& p_a \left(\xi\sigma^a \bar\xi \right)-\bm{\mu}  \ \approx \ 0  \,,  \label{const-sp-1}\\ [5pt]
\tilde l &=& p_a \left(\bar\pi\tilde\sigma^a \pi \right) -\bm{\mu} \ \approx \ 0  \,,  \label{const-sp-2}\\ [5pt]
u &=& N -\bar N \ \approx \ 0 \,,  \label{const-sp-3}
\end{eqnarray}
where
\begin{equation}
\label{N}
N=\xi^{\alpha}\pi_{\alpha}\,,\qquad \bar N=\bar\pi_{\dot\alpha} {\bar\xi}^{\dot\alpha}\,.
\end{equation}
In \p{D-constr}-\eqref{const-sp-3}, the quantities $p_a$,
$p_{\alpha}$, $\bar p^{\dot\alpha}$ are momenta for $x^a$,
$\theta^{\alpha}$, $\bar\theta_{\dot\alpha}$ respectively. The
canonical commutational relations are defined in terms of nonzero
Poisson brackets of the canonical variables:
\begin{equation}
\label{PB0}
\big\{ p_A , z^B \big\}={}-\delta^B_A\, :\qquad
\big\{ x^a, p_b \big\}=\delta^a_b\,,\quad
\big\{\theta^\alpha, p_{\beta} \big\}={}-{}\delta^\alpha_\beta\,,\quad
\big\{\bar\theta_{\dot\alpha}, \bar p^{\dot\beta} \big\}={}-{}\delta_{\dot\alpha}^{\dot\beta}\,.
\end{equation}
\begin{equation}
\label{DB-b-sp}
\big\{\xi^\alpha, \pi_\beta \big\}=\delta^\alpha_\beta\,,\qquad
\big\{\bar\xi^{\dot\alpha}, \bar\pi_{\dot\beta} \big\}=\delta^{\dot\alpha}_{\dot\beta}\,. \,\,\,
\end{equation}

Using \p{PB0} and \p{PB0}, we obtain that the non-vanishing Poisson
brackets of fermionic constraints \p{D-constr} and bosonic
constraints \eqref{const-sp}-\eqref{const-sp-3} have the form
\begin{equation}
\label{algebra-2}
\big\{ D_{\,\alpha}, \bar D_{\,\dot\alpha} \big\} \ = {} \ -2i \sigma^a_{\alpha\dot\alpha}\,p_a\,,
\end{equation}
\begin{equation}
\label{algebra-1}
\big\{ l, \tilde l \big\} \ ={} \ -K l_0\,,
\end{equation}
where
\be\lb{K}
K \ := \ N +\bar N
\ee
and $N$ and $\bar N$ are given by \p{N}.

Using the expressions \p{algebra-2} and \p{algebra-1}, we find that
all bosonic constraints \eqref{const-sp}-\eqref{const-sp-3} are the
first-class constraints, whereas four fermionic constraints
\p{D-constr} are the mixture of two first-class constraints and two
second-class constraints. The first-class constraints are
constructed on the basis of fermionic constraints by convolving them
with the matrix $\tilde p^{\dot\alpha\alpha}$. That is, the
constraints $\tilde p^{\dot\alpha\alpha}D_{\,\alpha}\approx 0$,
$\bar D_{\,\dot\alpha}\tilde p^{\dot\alpha\alpha}\approx 0$ are the
first-class ones which are the generators of local fermionic
$\kappa$-invariance. But these constraints are not independent due
to  the constraint \eqref{const-sp}.
It is impossible to extract independent
first- and second-class constraints from the constraints
\p{D-constr} in a covariant way without using
additional commuting variables.

However, a remarkable feature of the model under consideration is that it already contains the commuting spinor variables $\xi^{\alpha},\,\bar{\xi}^{\dot{\alpha}}$ by construction. This opens the opportunity of constructing independent fermionic first- and second-class constraints in a covariant way using these variables
(see, e.g., a similar consideration in \cite{FIL-2006}).
As a result, the odd constraints \p{D-constr} split into two second-class constraints and two first-class constraints as follows. The fermionic second-class constraints look like
\begin{equation}
\lb{sG-constr}
G:=\xi^{\alpha}D_{\,\alpha}\approx 0\,,
\qquad   \bar G:= \bar D_{\,\dot\alpha}\bar\xi^{\dot\alpha}\approx 0\,,
\qquad \bar G={}-(G)^*
\end{equation}
while the fermionic first-class constraints are
\begin{equation}\lb{fF-constr}
F:=\bar\xi_{\dot\alpha}\tilde p^{\dot\alpha\alpha}D_{\,\alpha}\approx 0\,,
\qquad   \bar F:= \bar D_{\,\dot\alpha}\tilde p^{\dot\alpha\alpha}\xi_{\alpha}\approx 0\,.
\qquad \bar F={}-(F)^*
\end{equation}
The algebra of constraints \p{sG-constr} and \p{fF-constr} has the form
\begin{equation}
\label{algebra-2-1}
\left\{ G, \bar G \right\} = {} -2i\, l -2i\bm{\mu}\,,\qquad
\left\{ F, \bar F \right\} = {}  2i (\xi p\,\bar\xi)\,l_0\,,
\end{equation}
\begin{equation}
\label{algebra-2a}
\left\{ G, \tilde l \right\} = F\,,\quad
\left\{ \bar G, \tilde l \right\} = \bar F\,, \qquad
\left\{ F, \tilde l \right\} = \pi^{\alpha}D_{\alpha}\,l_0\,,\quad
\left\{ \bar F, \tilde l \right\} = \bar D_{\dot\alpha}\bar\pi^{\dot\alpha}\,l_0\,,
\end{equation}
\begin{equation}
\label{algebra-2b}
\left\{ G, u \right\} = G\,,\quad
\left\{ \bar G, u \right\} ={} -\bar G\,, \qquad
\left\{ F, u \right\} ={} -F\,,\quad
\left\{ \bar F, u \right\} = \bar F\,.
\end{equation}
It is clearly seen that the only second-class constraints are $G$ and $\bar{G}$.

Let us describe  the main points of the procedure for extracting
first- and second-class constraints from fermionic constraints
\p{D-constr}. First, due to the constraints \eqref{const-sp} and
\eqref{const-sp-1}, \textit{i.\,e.} $p^2\approx 0$ and $\xi
p\bar\xi\approx \bm{\mu}\neq 0$, we have the identities \be
\label{delta}
\delta_{\alpha}^{\beta}=\frac{1}{\bm{\mu}}\left[\xi_{\alpha}(\bar\xi\tilde
p)^{\beta} +(p\bar\xi)_{\alpha}\xi^{\beta}\right],\qquad
\delta_{\dot\alpha}^{\dot\beta}=\frac{1}{\bm{\mu}}\left[\bar\xi_{\dot\alpha}(\tilde
p\xi)^{\dot\beta} +(\xi p)_{\dot\alpha}\bar\xi^{\dot\beta}\right].
\ee Using these identities, we present the constraints \p{D-constr}
in the form
$$
D_{\alpha}\approx\frac{1}{\bm{\mu}}\,\left[(p\bar\xi)_{\alpha}G+\xi_{\alpha}F\right]\,,\qquad
\bar D_{\dot\alpha}\approx\frac{1}{\bm{\mu}}\,\left[(\xi p)_{\dot\alpha}\bar G+\bar\xi_{\dot\alpha}\bar F\right]\,.
$$
We see that the constraints $G$  and $F$ are equivalent to $D_{\alpha}$ and
the constraints $\bar G$  and $\bar F$ are equivalent to $\bar D_{\dot\alpha}$.

The canonical Hamiltonian in the considered system is defined as a linear combination of all first-class constraints:
\begin{equation}
\label{H-sp}
H_{0} \ = \ \frac12\,e\, l_0 \  - \
\lambda\, l \  - \  \tilde\lambda\, \tilde l\  - \  iA u \  + \ \chi F \  + \  \bar\chi\bar F \,.
\end{equation}
where the constraints $e$, $\lambda$, $\tilde\lambda$, $u$, $\chi$ and $\bar\chi$ are
the Lagrange multipliers.

\setcounter{equation}{0}

\section{The continuous spin superparticle in curved superspace: problem of Lagrangian and constraints}

We begin to consider the superparticle in a curved superspace parameterized by supercoordinates
\be
z^M=(x^m,\theta^{\mu},\bar\theta_{\dot\mu})\,, \qquad \mbox{where} \qquad
x^m=(x^m)^*\,, \quad \bar\theta^{\dot\mu}=(\theta^{\mu})^*\,, \quad \bar\theta^{\dot\mu}=\epsilon^{\dot\mu\dot\nu}\bar\theta_{\dot\nu}\,.
\ee
Also, the infinite spin particle is characterized by additional target space coordinates which are the commuting Weyl spinors $\xi^{\alpha}$, $\bar\xi^{\dot\alpha}$, $\pi_{\alpha}$, $\bar\pi_{\dot\alpha}$ with flat indices like in the previous section.

As a first step, we perform a direct curved superspace
generalization of the Lagrangian \p{L-super-1}. For this, we make the
following  replacements in \p{L-super-1} :
\begin{equation}
\label{repl-curl}
\omega^a \quad \rightarrow \quad \dot z^M E_M{}^a\,,\qquad
\dot\xi^{\alpha} \quad \rightarrow \quad \dot\xi^{\alpha}+\dot z^M \Omega_{M}{}^{\alpha\beta}\xi_{\beta}\,,
\qquad
\dot{\bar\xi}^{\dot\alpha} \quad \rightarrow \quad
\dot{\bar\xi}^{\dot\alpha}+\dot z^M \Omega_{M}{}^{\dot\alpha\dot\beta}{\bar\xi}_{\dot\beta}\,,
\end{equation}
where $E_M{}^A(z)$ is the supervielbein matrix and
$\Omega_{M}{}^{\alpha\beta}(z)$,
$\Omega_{M}{}^{\dot\alpha\dot\beta}(z)$ are the components of spin
superconnection (see, e.g., \cite{Ideas}). Let us turn attention to
the rules of contraction in the superindex $M$ in \p{repl-curl}: $A^M
B_M =A^m B_m +A^\mu B_\mu +A_{\dot\mu} B^{\dot\mu}$. For example,
$$
\dot z^M E_M{}^a=\dot x^m E_m{}^a+\dot \theta^\mu E_\mu{}^a+\dot {\bar\theta}_{\dot\mu} E^{\dot\mu}{}^a\,.
$$
Thus, at this stage the curved superspace generalization of  the
supersymmetric-invariant Lagrangian \p{L-super-1} has the form
\begin{equation}
\label{L-super-curl}
L \ = \ \frac{1}{2e}\,\mathscr{W}^{\,a} \mathscr{W}_{a}
\ +  \ \mathcal{D}\xi^{\alpha} \pi_{\alpha}  \   +  \ \bar\pi_{\dot\alpha} \mathcal{D}{\bar\xi}^{\dot\alpha}
\  - \  (\lambda+\tilde\lambda)\bm{\mu}\,,
\end{equation}
where now
(see \p{Omega-1} for comparison)
\begin{equation}
\label{Omega-curl}
\mathscr{W}^{\,a} \ := \ \dot z^M E_M{}^a + \lambda (\xi\sigma^a \bar\xi ) + \tilde\lambda (\bar\pi\tilde\sigma^a \pi )
\end{equation}
and (see \p{nabla-xi} for comparison)
\begin{equation}
\label{cD-xi}
\mathcal{D}\xi^{\alpha}:=\dot \xi^{\alpha}+\dot z^M \Omega_{M}{}^{\alpha\beta}\xi_{\beta} +iA\xi^{\alpha}\,,\qquad
\mathcal{D}{\bar\xi}^{\dot\alpha}:=\dot {\bar\xi}^{\dot\alpha}+\dot z^M \Omega_{M}{}^{\dot\alpha\dot\beta}{\bar\xi}_{\dot\beta}-iA{\bar\xi}^{\dot\alpha}\,.
\end{equation}

In the flat case, when the vielbein $E_M{}^A(z)$ has only the following
nonzero components \footnote{ Here
$\theta_{\mu}=\epsilon_{\mu\nu}\theta^{\nu}$ and
$(\tilde\sigma^a)^{\dot\mu\mu}=\epsilon^{\mu\nu}\epsilon^{\dot\mu\dot\nu}(\sigma^a)_{\nu\dot\nu}$}
$$
\overset{_{\mathrm{\ o}}}{E}_M{}^A=\left(
\overset{_{\mathrm{\ o}}}{E}_m{}^a=\delta_m^a\,,\
\overset{_{\mathrm{\ o}}}{E}_\mu{}^\alpha=\delta_\mu^\alpha\,,\
\overset{_{\mathrm{\ o}}}{E}{}^{\dot\mu}{}_{\dot\alpha}=\delta^{\dot\mu}_{\dot\alpha}\,,\
\overset{_{\mathrm{\ o}}}{E}_{\mu}{}^a={}-{}i(\sigma^a)_{\mu\dot\mu}\bar\theta^{\dot\mu}\,,\
\overset{_{\mathrm{\ o}}}{E}{}^{\dot\mu}{}^a={}-{}i(\tilde\sigma^a)^{\dot\mu\mu}\theta_{\mu}
\right)
$$
and the spin connection is zero, the Lagrangian \p{L-super-curl} becomes
the flat space Lagrangian  \p{L-super-1}.

To construct a canonical formulation, one introduces the
supermomenta $p_M$ corresponding to the supercoordinates $z^M$:
\begin{equation}
\label{mom-z-def}
p_M=\frac{\partial L}{\partial \dot z^M}\,:\qquad p_m=\frac{\partial L}{\partial \dot x^m}\,,\quad
p_\mu=\frac{\partial L}{\partial \dot \theta^\mu}\,,\quad \bar p^{\dot\mu}=\frac{\partial L}{\partial \dot{\bar\theta}_{\dot\mu}}\,.
\end{equation}
Relations \p{mom-z-def} yield
\begin{equation}
\label{mom-z}
p_M \ = \ \frac{1}{e}E_M{}^a \mathscr{W}_a \ + \ \Omega_{M}{}^{\alpha\beta}\pi_{\alpha}\xi_{\beta}
 \ + \ \Omega_{M}{}^{\dot\alpha\dot\beta}\bar\pi_{\dot\alpha}\bar\xi_{\dot\beta}\,.
\end{equation}
The canonical Poisson brackets of the supermomenta $p_M$ and the
supercoordinates $z^M$ look like (see \p{PB0}) \footnote{For
conveninece we use the notation for the momentum corresponding to
$\bar\theta_{\dot\mu}$  with a bar, that is $\bar p^{\dot\mu}$.}
\begin{equation}
\label{PBc}
\left\{ p_M, z^N \right\}={}-{}\delta^N_M\, :\qquad
\left\{ p_m, x^n, \right\}={}-{}\delta^n_m\,,\quad
\left\{p_{\mu} ,\theta^\nu\right\}={}-{}\delta^\nu_\mu\,,\quad
\left\{\bar p^{\dot\mu} ,\bar\theta_{\dot\nu}\right\}={}-{}\delta^{\dot\mu}_{\dot\nu}\,.
\end{equation}
The commuting Weyl spinors $\xi^\alpha$, $\bar\xi^{\dot\alpha}$ and
$\pi_\alpha$, $\bar\pi_{\dot\alpha}$ form canonical pairs:
they nonvanishing Poisson brackets have the form \p{DB-b-sp},
i.e.
\begin{equation}
\label{DB-b-sp1}
\left\{\pi_\beta ,\xi^\alpha \right\}_{{}_D}={}-{}\delta^\alpha_\beta\,,\qquad
\left\{\bar\pi_{\dot\beta} ,\bar\xi^{\dot\alpha} \right\}_{{}_D}={}-{}\delta^{\dot\alpha}_{\dot\beta}\,.
\end{equation}

With respect to complex conjugation, the momenta transform as follows:
\be\lb{cc-mom}
(p_m)^*=p_m\,,\qquad
(p_{\mu} )^*= \bar p_{\dot\mu}=\epsilon_{\dot\mu\dot\nu}\bar p^{\dot\nu}\,,\qquad
(\pi_\alpha)^*=\bar\pi_{\dot\alpha} \,.
\ee

It is convenient to represent expressions \p{mom-z} in the form:
\begin{equation}
\label{c-mom-z}
\mathcal{P}_M= \frac{1}{e}E_M{}^a \mathscr{W}_a\,,
\end{equation}
where
\begin{equation}
\label{c-mom-def}
\mathcal{P}_M:=p_M-\Omega_{M}{}^{\alpha\beta}\pi_{\alpha}\xi_{\beta}
-\Omega_{M}{}^{\dot\alpha\dot\beta}\bar\pi_{\dot\alpha}\bar\xi_{\dot\beta}\,.
\end{equation}
We rewrite the $\mathcal{P}_M$ as follows:
\begin{equation}
\label{c-mom-M}
\mathcal{P}_M\ =\ p_M \ + \ \frac12\,\Omega_{M}{}^{ab}M_{ab}\,,
\end{equation}
where
\begin{equation}
\label{Not-sp}
M_{ab} = \xi\sigma_{ab}\pi - \bar\pi\tilde\sigma_{ab}\bar\xi \,,\qquad M_{ab} =(M_{ab})^*
\end{equation}
are the angular momenta generators forming the Lorentz algebra
\begin{equation}
\label{M-alg}
\big\{ M_{ab} ,M_{cd} \big\} \ = \ \eta_{ac}M_{bd}+\eta_{bd}M_{ac}-\eta_{ad}M_{bc}-\eta_{bc}M_{ad}\,.
\end{equation}
The superconnections with vector and spinor indices relate to each
other by:
\begin{equation}
\label{omega}
\Omega_M{}^{ab} = (\sigma^{ab})_{\alpha\beta}\Omega_{M}{}^{\alpha\beta} - (\tilde{\sigma}^{ab})_{\dot\alpha\dot\beta}\Omega_{M}{}^{\dot\alpha\dot\beta}\,,
\end{equation}
\begin{equation}
\label{omega1}
\Omega_{M}{}^{\alpha\beta}=\frac{1}{2}\,\Omega_M{}^{ab} (\sigma_{ab})^{\alpha\beta} \,,\qquad
\Omega_{M}{}^{\dot\alpha\dot\beta}=-\frac{1}{2}\,\Omega_M{}^{ab}(\tilde{\sigma}_{ab})^{\dot\alpha\dot\beta}\,.
\end{equation}
The Poisson brackets of quantities \p{c-mom-M} look like
\begin{equation}
\label{cal-P-alg}
\big\{ \mathcal{P}_M ,\mathcal{P}_N \big\}_{{}_D} \ ={} \ -\frac{1}{2}\,R_{MN}{}^{ab}M_{ab}\,,
\end{equation}
where \footnote{
Here, as usual, $M$ and $N$ in the exponent are equal to $0$ or $1$ for vector and spinor indices, respectively.
}
\begin{equation}
\label{R-expr}
R_{MN}{}^{ab} \ = \ \partial_M\Omega_N{}^{ab} \ - \ (-)^{MN}\partial_N\Omega_M{}^{ab}
\ + \ \Omega_M{}^{ac}\Omega_N{}_c{}^{b} \ - \ (-)^{MN}\Omega_N{}^{ac}\Omega_M{}_c{}^{b}
\end{equation}
is the superspace curvature tensor.

Introducing the inverse supervielbein
$$
E_A{}^M(z)\,,\qquad E_M{}^AE_A{}^N=\delta_M^N\,,\quad E_A{}^ME_M{}^B=\delta_A^B\,,
$$
we rewrite the quantities \p{c-mom-M} with world superindices
through the quantities with tangent (flat) superindices:
\begin{equation}
\label{c-mom-t-def}
\mathcal{P}_A \ = \ E_A{}^M\mathcal{P}_M\ =\ p_A \ + \ \frac12\,\Omega_{A}{}^{ab}M_{ab}\,,
\end{equation}
where
\begin{equation}
\label{mom-t-def}
p_A=E_A{}^M p_M\,,
\qquad
\Omega_{A}{}^{ab} \ = \ E_A{}^M\Omega_{M}{}^{ab}\,.
\end{equation}
The Poisson brackets of the quantities \p{c-mom-t-def} are
\footnote{
Compared to the notation in our paper \cite{BFIK-24-1},
here the definition of the `angolonomy coefficients' $C_{AB}{}^{C}$ and supertorsion components $T_{AB}{}^{C}$
contain an additional factor of 2, as is customary in the literature on supersymmetry (see \cite{WessBagger,Ideas}).
In addition, we use an additional sign ``$-$'' in the definition of  the `angolonomy coefficients' $C_{AB}{}^{C}$ and supertorsion components $T_{AB}{}^{C}$. As result, our definitions coincide with the definitions in \cite{Ideas}
(in  \cite{WessBagger}, the definition of supertorsion differs by the sign ``$-$'' compared to the definition in  \cite{Ideas}).
}
\begin{equation}
\label{cal-Pa-alg}
\big\{ \mathcal{P}_A ,\mathcal{P}_B \big\}_{{}_D} \ = \ {} -\frac{1}{2}\,R_{AB}{}^{cd}M_{cd} \ - \ C_{AB}{}^{C}\mathcal{P}_C\,,
\end{equation}
where
\begin{eqnarray}
\label{R-expr-L}
R_{AB}{}^{cd} & = & (-)^{M(N+B)}E_A{}^M E_B{}^N R_{MN}{}^{cd} \\ [6pt]
\nonumber
& = & E_{A}{}^N \partial_N \Omega_{B}{}^{cd} - (-)^{AB}E_{B}{}^N \partial_N \Omega_{A}{}^{cd} +
\Omega_{A}{}^{ch}\Omega_{B}{}_h{}^{d}- (-)^{AB}\Omega_{B}{}^{ch}\Omega_{A}{}_h{}^{d}-
C_{AB}{}^D\Omega_{D}{}^{cd}
\end{eqnarray}
is the target space curvature tensor and
\begin{equation}
\label{Tt-expr}
C_{AB}{}^C  = {}  \Big( E_{A}{}^N \partial_N E_{B}{}^M - (-)^{AB}E_{B}{}^N \partial_N E_{A}{}^M\Big)\,E_M{}^C
\end{equation}
is the connection-free part of the torsion. Note that
\begin{equation}
\label{C-C}
C_{AB}{}^{C}  =  (-)^{M(N+B)}E_A{}^M E_B{}^N C_{MN}{}^{C}\,,\quad
\mbox{where}\quad C_{MN}{}^A  ={} - \Big( \partial_{M} E_{N}{}^A - (-)^{MN}\partial_{N} E_{M}{}^A\Big)\,.
\end{equation}
Besides, below we will use the identity \cite{Ideas}
\begin{equation}
\label{R-R-sp}
\frac12\,R_{AB}{}^{cd} M_{cd}\ =\ R_{AB}{}^{\gamma\delta} M_{\gamma\delta} +
R_{AB}{}^{\dot\gamma\dot\delta}\bar M_{\dot\gamma\dot\delta}\,,
\end{equation}
where, similarly to \p{omega}, the spin-tensor components of the
angular momentum tensor \p{Not-sp}  defined by
\begin{equation}
\label{M-sp-s}
M^{ab} = (\sigma^{ab})_{\alpha\beta}{M}^{\alpha\beta} - (\tilde{\sigma}^{ab})_{\dot\alpha\dot\beta}\bar{M}^{\dot\alpha\dot\beta}\,,
\end{equation}
are equal to
\begin{equation}
\label{M-sp-comp}
{M}^{\alpha\beta}={} -\xi^{(\alpha}\pi^{\beta)}\,,\qquad \bar{M}^{\dot\alpha\dot\beta}={} -\bar\xi^{(\dot\alpha}\bar\pi^{\dot\beta)}
 \,,\qquad \bar{M}^{\dot\alpha\dot\beta} =({M}^{\alpha\beta})^*\,.
\end{equation}

In terms of the quantities \p{c-mom-t-def}, expressions
\p{c-mom-z} take the form
\be \label{c-mom-a}
\mathcal{P}_a\ = \
\frac{1}{e}\,\mathscr{W}_a\,,
\ee
\be \label{c-mom-al}
\mathcal{P}_\alpha \ =\  0\,, \qquad \bar{\mathcal{P}}^{\dot\alpha}
\ =\  0\,.
\ee
Similarly to \p{cc-mom}, the covariant momenta
transform under complex conjugation as follows \be
(\mathcal{P}_a)^*=\mathcal{P}_a\,,\qquad
(\mathcal{P}_\alpha)^*=\bar{\mathcal{P}}_{\dot\alpha}=\epsilon_{\dot\alpha\dot\beta}\bar{\mathcal{P}}^{\dot\beta}
\,. \ee

Using the Lagrangian \p{L-super-curl}, we obtain the canonical Hamiltonian in the form (see \p{H-sp} for comparison)
\begin{equation}
\label{H-sp-curv}
H_{C} \ = \ \frac12\,e \ell_0 \  - \
\lambda \ell \  - \  \tilde\lambda \tilde\ell\  - \  iA u\,,
\end{equation}
where \p{const-sp}-\p{const-sp-3}
\begin{eqnarray}
\ell_0 &:=& \mathcal{P}_a \mathcal{P}^a  \,, \label{const-sp-curv}\\ [5pt]
\ell &:=& \mathcal{P}_a \left(\xi\sigma^a \bar\xi \right)-\bm{\mu}  \,,  \label{const-sp-1-curv}\\ [5pt]
\tilde\ell &:=& \mathcal{P}_a \left(\bar\pi\tilde\sigma^a \pi \right) -\bm{\mu}  \,,  \label{const-sp-2-curv}\\ [5pt]
u &:=& N -\bar N \,.  \label{const-sp-3-curv}
\end{eqnarray}
In \p{H-sp-curv}, the quantities $e$, $\lambda$, $\tilde\lambda$ and $A$ are the Lagrange multipliers
(the procedure for processing them and their momenta is exactly the same as for the flat space superparticle considered in the previous section).
It is evident that
 the quantities \p{const-sp-curv}-\p{const-sp-3-curv} define the bosonic constraints
\be\lb{cov-mom-even}
\ell_0\approx0\,, \qquad \ell\approx0\,, \qquad  \tilde\ell\approx0\,, \qquad u\approx0\,.
\ee

In addition to these bosonic constraints, the definition of spinor momenta gives us fermionic constraints
\p{c-mom-al}:
\be\lb{cov-mom-odd}
\mathscr{D}_{\alpha} \ := \ i\,\mathcal{P}_{\alpha} \ \approx \ 0\,, \qquad
\bar{\mathscr{D}}^{\dot\alpha} \ := \ i\,\bar{\mathcal{P}}^{\dot\alpha} \ \approx \ 0\,.
\ee

In the flat case, when inverse vielbein $E_A{}^M(z)$ has only the following nonzero components
$$
\overset{_{\mathrm{\ o}}}{E}_A{}^M=\left(
\overset{_{\mathrm{\ o}}}{E}_a{}^m=\delta^m_a\,,\
\overset{_{\mathrm{\ o}}}{E}_\alpha{}^\mu=\delta^\mu_\alpha\,,\
\overset{_{\mathrm{\ o}}}{E}{}^{\dot\alpha}{}_{\dot\mu}=\delta_{\dot\mu}^{\dot\alpha}\,,\
\overset{_{\mathrm{\ o}}}{E}_\alpha{}^m= i (\sigma^m)_{\alpha\dot\alpha}\bar\theta^{\dot\alpha}\,,\
\overset{_{\mathrm{\ o}}}{E}{}^{\dot\alpha}{}^m= i (\tilde\sigma^m)^{\dot\alpha\alpha}\theta_{\alpha}
\right),
$$
spin connection vanishes and $\overset{_{\mathrm{\ o}}}{\mathcal{P}}_m=p_m$, $\overset{_{\mathrm{\ o}}
}{\mathcal{P}}_\mu=p_\mu$,
$\overset{_{\mathrm{\ o}}}{\mathcal{P}}_{\dot\mu}=p_{\dot\mu}$. Thus, in the flat limit  the equalities
$\overset{_{\mathrm{\ o}}}{\mathcal{P}}_a=p_a$, $\overset{_{\mathrm{\ o}}}{\mathcal{P}}_\alpha={}-iD_\alpha$,
$\overset{_{\mathrm{\ o}}}{\mathcal{P}}_{\dot\alpha}=\epsilon_{\dot\alpha\dot\beta}\overset{_{\mathrm{\ o}}}
{\mathcal{P}}^{\dot\beta}=-i\bar D_{\dot\alpha}$ take place and
the constraints \p{cov-mom-odd} become the flat fermionic constraints \p{D-constr}.

To sum up the interim results, we have arrived at the Hamiltonian of superparticle in curved superspace \p{H-sp-curv} on the basis of Lagrangian
\p{L-super-curl} and found the system of corresponding bosonic \p{cov-mom-even} and fermionic \p{cov-mom-odd}
constraints. Since the Lagrangian was obtained purely formally only by supercovariantizating the flat superspace Lagrangian,
we cannot state that the Lagrangian \p{L-super-curl} is a correct Lagrangian for the infinite spin superparticle in curved superspace.
The matter is that the formal covariantization does not take into account possible non-minimal terms which
can be present in true Lagrangian. Therefore, the constraints obtained on the basis of Lagrangian \p{L-super-curl}
cannot be considered as correct constraints for the superparticle in curved superspace.
In particular, it is unclear to which classes the constraints given by relations \p{cov-mom-even} and
\p{cov-mom-odd} belong in the correct theory.
This should be  extremely important for a complete description of the canonical structure of the model under
consideration.
Thus, our next stage is to establish a correct form of the first- and second-class constraints
in our model  without changing the total number of constraints.
Taking into account the results on purely bosonic infinite spin particle \cite{BFIK-24-1},
one can expect that the bosonic constraints \p{cov-mom-even} should be modified in some way
by supergeometry dependent terms to get a system of correct first-class constraints.
Besides, one can expect some restrictions on supergeometry.
As to fermionic constraints \p{cov-mom-odd}, they should be decomposed
in an explicit supercovariant form into
first- and second-class constraints as it was done in the previous section.
That is, the following requirements must be fulfilled:
\begin{enumerate}
\item
The bosonic constraints  \p{cov-mom-even} must be modified to get the first-class constraints.
\item
The fermionic constraints \p{cov-mom-odd} must be decomposed into first- and second-class constraints.
\item
No additional constraints in the curved superspace should arise.
\end{enumerate}
The realization of these requirements will be considered in the next sections.\footnote{Note that the modifications of the constraints in fact mean a modification of the Lagrangian \p{L-super-curl}.}

\setcounter{equation}{0}

\section{Curved background superspace
 construction of the Lagrangian and constraints}

\subsection{Construction of the bosonic first-class constraints}

In this subsection we will discuss a modification of the bosonic
constraints to get a system of first-class constraints. Let us begin
with calculating the Poisson brackets of the constraints  $\ell$
\p{const-sp-1-curv} and $\tilde\ell$ \p{const-sp-2-curv}. The direct
calculations lead to
\be \label{PB-llp}
\big\{ \ell ,\tilde \ell
\big\}  \ = {} \ - K \ell_0^\prime \  - \ \frac{1}{4}
\left(\xi\sigma^{ab} \pi +\bar\pi\tilde\sigma^{ab} \bar\xi
\right)R_{ab}{}^{cd} M_{cd}\,u \ - \ \left(\xi\sigma^a \bar\xi
\right)\left(\bar\pi\tilde\sigma^b \pi \right) T_{ab}{}^C
\mathcal{P}_C \,,
\ee
where we have used the relation \footnote{These
relations are derived with the help of the identity
$(\sigma^{[a})_{\alpha\dot{\alpha}}(\tilde{\sigma}^{b}])_{\dot{\beta}\beta}
=
-(\sigma^{ab})_{\alpha}{}^{\gamma}\epsilon_{\gamma\beta}\epsilon_{\dot{\alpha}\dot{\beta}}-
\epsilon_{\dot{\alpha}\dot{\gamma}}(\tilde{\sigma}^{ab})^{\dot{\gamma}}{}_{\dot{\beta}}\epsilon_{\alpha\beta}$}
\bea \label{ident}
\left(\xi\sigma^{[a} \bar\xi
\right)\left(\bar\pi\tilde\sigma^{b]} \pi \right)  \ = \  {}
-\frac12\,K\,M^{ab} + \frac12\,u \left(\xi\sigma^{ab} \pi
+\bar\pi\tilde\sigma^{ab} \bar\xi \right),
\eea
and the following
notation has been introduced:
\begin{equation}
\label{K-def}
K \ := \ N \  + \  \bar N \,,
\end{equation}
\be
\label{L0-p}
\ell_0^\prime  \ := \
\ell_0 \  - \ \frac{1}{4}\,R_{ab}{}^{cd} M^{ab} M_{cd}\,.
\ee
The quantity $u$ is defined in \p{const-sp-3-curv} and
\begin{equation}
\label{T-expr}
T_{AB}{}^C \ = \ C_{AB}{}^C \ + \ \Omega_{AB}{}^C \ - \ (-)^{AB}\Omega_{BA}{}^C \ = \ (-)^{M(N+B)}E_A{}^M E_B{}^N T_{MN}{}^{C}
\end{equation}
is the supertorsion. Note that $T_{ab}{}^{\alpha}=C_{ab}{}^{\alpha}$
and $T_{ab}{}^{\dot\alpha}=C_{ab}{}^{\dot\alpha}$. We see that
relation \p{PB-llp} does not form the first-class algebra.

Let us begin with modifications of the constraints. We considers the
quantity $\ell_0^\prime\approx0$ as the first step modification
$\ell_0 \to \ell_0^\prime$ of the constraint $\ell_0\approx0$ and
check a Poisson bracket of this new constraint with the constraints
$\ell\approx0$ and $\tilde{\ell}\approx0$.

So let us consider the Poisson brackets of the quantity \p{L0-p} with $\ell$ and $\tilde \ell$. The result of calculations has the form
\be\label{PB-1F0-l}
\big\{ \ell_0^\prime \, , \, \ell \big\} \ = \
2\left(\xi\sigma^a \bar\xi \right)\mathcal{P}^b R_{ab}{}^{cd} M_{cd}
\ - \ \frac{1}{4}\left(\xi\sigma^h \bar\xi \right)\mathcal{D}_h {R}_{ab}{}^{cd}M^{ab}M_{cd}
\  + \
2\left(\xi\sigma^a \bar\xi \right)\mathcal{P}^b T_{ab}{}^C \mathcal{P}_C\,,
\ee
\be
\label{PB-1F0-tl}
\big\{\ell_0^\prime \, , \, \tilde \ell \big\} \ = \
2\left(\bar\pi\tilde\sigma^a \pi \right)\mathcal{P}^bR_{ab}{}^{cd} M_{cd}
\ - \ \frac{1}{4}\left(\bar\pi\tilde\sigma^h \pi \right)\mathcal{D}_h {R}_{ab}{}^{cd}M^{ab}M_{cd}
\  + \
2\left(\bar\pi\tilde\sigma^a \pi \right)\mathcal{P}^b T_{ab}{}^C \mathcal{P}_C\,,
\ee
where
\be
\label{D-cur}
\mathcal{D}_h {R}_{ab}{}^{cd} =
\partial_h {R}_{ab}{}^{cd}
+\Omega_{h\,a}{}^g{R}_{gb}{}^{cd} +\Omega_{h\,b}{}^g{R}_{ag}{}^{cd}
+\Omega_{h\,}{}^c{}_g{R}_{ab}{}^{gd}
+\Omega_{h\,}{}^d{}_g{R}_{ab}{}^{cg}
\ee
and $\partial_a = E_a{}^{M}\partial_M$.
We see that the quantities $\ell_0^\prime$, $\ell$ and $\tilde \ell$
defined in \p{L0-p}, \p{const-sp-1-curv} and \p{const-sp-2-curv}  do not form a closed algebra in the general case.
To obtain a closed algebra, we should impose some restrictions on supergeometry and perhaps make further
modifications of the constraints.

As we pointed out, the modifications of the constraints should not
change the number of first-class bosonic constraints. Let us pay
attention that the initial constraints $\ell$ \p{const-sp-1-curv}
and $\tilde \ell$ \p{const-sp-2-curv} contain the terms that are linear
in  $\mathcal{P}_a$. Modifying these constraints to include higher
powers of $\mathcal{P}_a$ will almost inevitably lead to an increase
in the number of constraints since the algebra of the constraints
\p{PB-1F0-l},\,\p{PB-1F0-tl} cannot be closed only on the
constraints $\ell_0$, $\ell$, $\tilde{\ell}$. To avoid the emergence
of additional constrains, we should preserve linear dependence of
expressions on $\mathcal{P}_a$. But $\ell$ and $\tilde \ell$ depend
on $\mathcal{P}_a$ through the expressions $\mathcal{P}_a
\left(\xi\sigma^a \bar\xi \right)$ and $\mathcal{P}_a
\left(\bar\pi\tilde\sigma^a \pi \right)$. To preserve only such linear in $\mathcal{P}_a$
expressions at
modifications of the constraints, we are forced to impose  restrictions on the background
supergeometry.

Let us consider the right-hand sides of relations \p{PB-llp},
\p{PB-1F0-l} and \p{PB-1F0-tl} and find out under what conditions
these relations will be closed when the constraints
$\ell^{\prime}_{0},\, \ell,\, \tilde{\ell}$ are satisfied. Taking
into account the equalities \be \left(\xi\sigma^a \bar\xi
\right)M_{ab}={}-\frac12\left(\xi\sigma_b \bar\xi \right)K\,,\qquad
\left(\bar\pi\tilde\sigma^a \pi
\right)M_{ab}=\frac12\left(\bar\pi\tilde\sigma_b \pi \right)K\,,
 \ee
we see that the needed terms  $\mathcal{P}_a
\left(\xi\sigma^a \bar\xi \right)$ and $\mathcal{P}_a
\left(\bar\pi\tilde\sigma^a \pi \right)$ will appear on the
right-hand sides of \p{PB-llp}, \p{PB-1F0-l} and \p{PB-1F0-tl} only in the case when
the components ${R}_{ab}{}^{cd}$ of the curvature tensor of background geometry are taken in the form
\be
\label{R-dSitter}
{R}_{ab}{}^{cd} \ ={} \ k
\left(\delta_a^c\delta_b^d-\delta_a^d\delta_b^c\right),
\ee
where $k$ is a constant. As is well know, relation
\p{R-dSitter} defines the constant curvature space-time. At zero,
positive or negative $k$ we get the Minkowski, de Sitter, and
anti-de Sitter spaces, respectively.

As is well known, only the anti-de Sitter space is consistent with conventional supersymmetry \cite{IvSor} (also see, e.g., \cite{Ideas})
\footnote{The formulation of supersymmetry models on de Sitter space requires some modification of
the superalgebra (see, e.g., \cite{PNS,AFM,BFKP,Kuz} and references therein.}.
Therefore, below we will consider only the anti-de Sitter space with
\be
k ={} -\varkappa^2 \,,
\ee
where $\varkappa$ is a real number.
In this case nonvanishing components of the curvature and the torsion tensors take the form \cite{IvSor,Ideas}:
\be \lb{deS-curv}
\begin{array}{c}
R_{ab}{}^{cd} ={} - 2\varkappa^2 \delta_{[a}^c\delta_{b]}^d \,,\qquad
R_{\alpha\beta}{}^{\gamma\delta}
={} -4\varkappa\,\delta_{(\alpha}^\gamma\delta_{\beta)}^\delta \,,\qquad
R_{\dot\alpha\dot\beta}{}^{\dot\gamma\dot\delta}
={} 4 \varkappa\,\delta_{(\dot\alpha}^{\dot\gamma}\delta_{\dot\beta)}^{\dot\delta}
\,, \\ [6pt]
\displaystyle T_{\alpha\dot\beta}{}^{c}={}-2i(\sigma^c)_{\alpha\dot\beta} \,,\qquad
T_{a\beta}{}^{\dot\gamma}={}\frac{i}{2}\,\varkappa\,(\sigma_a)_{\beta}^{\dot\gamma} \,,\qquad
T_{a\dot\beta}{}^{\gamma}={}-\frac{i}{2}\,\varkappa\,(\sigma_a)_{\dot\beta}^{\gamma} \,.
\end{array}
\ee
These quantities satisfy the reality conditions:
\be
(R_{ab}{}^{cd})^*=R_{ab}{}^{cd}\,,\quad
(R_{\alpha\beta}{}^{\gamma\delta})^*={}-R_{\dot\alpha\dot\beta}{}^{\dot\gamma\dot\delta}\,,\qquad
(T_{\alpha\dot\beta}{}^{c})^*={}-T_{\beta\dot\alpha}{}^{c}\,,\quad (T_{a\beta}{}^{\dot\gamma})^*={}-T_{a\dot\beta}{}^{\gamma}\,.
\ee

We emphasize that in the curved superspace the
equalities $T_{\alpha\dot\beta}{}^{c}=C_{\alpha\dot\beta}{}^{c}$, $T_{a\beta}{}^{\dot\gamma}=C_{a\beta}{}^{\dot\gamma}$,
$T_{a\dot\beta}{}^{\gamma}=C_{a\dot\beta}{}^{\gamma}$ hold,
which reflects the fact that there are no such corresponding nonvanishing components of spin connections.
Besides, the curvature tensor \p{R-dSitter} of
constant curvature spaces is a covariantly constant:
 \be
\label{zero} \mathcal{D}_h {R}_{ab}{}^{cd} = 0.
 \ee

As result, in anti-de Sitter superspace \p{deS-curv} we get the following Poisson bracket of the constraints $\ell$ and $\tilde{\ell}$
\be \label{PB-ll-T000}
\big\{ \ell \,,\,\tilde \ell \big\} ={} - K \ell_0^{\,\prime\prime}\,,
\ee
where we have used the relations
\be
\label{RMM-dS}
-  \frac{1}{4}\,{R}_{ab}{}^{cd}
M^{ab}M_{cd}=\frac{1}{2}\,k\left(N^2+ \bar N^2 \right),\qquad
-  \frac{1}{4}\left(\xi\sigma^{ab} \pi +\bar\pi\tilde\sigma^{ab} \bar\xi \right)R_{ab}{}^{cd} M_{cd}
=\frac{1}{2}\,k\,K\,u\, ,
\ee
and the second stage modified initial constraint has been introduced
\be \lb{Lti2}
\ell_0^{\,\prime\prime} \ := \ \ell_0^{\prime} -\frac{1}{2}\,k\,u^2 \ = \
\ell_0-  \frac{1}{4}\,{R}_{ab}{}^{cd}M^{ab}M_{cd}  -\frac{1}{2}\,k\,u^2 \ = \
\ell_0+kN\bar N \ = \
\mathcal{P}_a \mathcal{P}^a+kN\bar N\,.
\ee

Let us consider the Poisson bracket of the constraint $\ell_0^{\,\prime\prime}$ \p{Lti2} with the constraints
\p{const-sp-1-curv}, \p{const-sp-2-curv}. The direct calculations lead to the following result:
\be \label{PB-ll-T00}
\big\{ \ell_0^{\,\prime\prime}\, ,\, \ell \big\} ={} -2k K(\ell+\bm{\mu}) \,,
\qquad
\big\{ \ell_0^{\,\prime\prime} \, , \, \tilde \ell \big\} ={} 2k K(\tilde \ell+\bm{\mu}) \,.
\ee

This algebra is not closed due to the presence of $\bm{\mu}$-dependent terms in the right-hand side of \p{PB-ll-T00}.
Therefore, the bosonic constraints still need to be modified in curved superspace.
This modification can be made exactly as in the purely bosonic case \cite{BFIK-24-1}
by adding extra terms with operators $N$ and $\bar N$.

Let us begin with the modification of the constraint \p{Lti2} requiring that the modified constraint commutes with
$\ell$ and $\tilde \ell$. Using the Poisson brackets
\be
\left\{ N,\ell \right\} =\left\{ \bar N,\ell \right\} ={}-(\ell+\bm{\mu})\,,\qquad
\left\{ N,\tilde \ell \right\} =\left\{ \bar N,\tilde \ell \right\} =\tilde \ell+\bm{\mu}
\ee
and \p{PB-ll-T00}, we see that the quantity, which is equal
to the sum of $\ell_0^{\prime\prime}$ plus the additional term
\be\lb{add-terms}
-\frac{k}{2}\,K^2 \qquad\quad \mbox{or}\qquad\quad -2{k}N\bar N \qquad\quad \mbox{or}\qquad\quad -{k}(N^2+\bar N^2)\,,
\ee
has zero Poisson brackets with $\ell$ and $\tilde \ell$.
However, at the constraint $\mathcal{U} = N -\bar N \approx0$,
the following equalities take place:
\be
-\frac{k}{2}\,K^2 = -2{k}N\bar N = -{k}(N^2+\bar N^2)\,,
\ee
which allows us to reduce the number of variants in \p{add-terms}. We choose the second option and define the quantity
\be\lb{L0-a}
\bm{\ell}_{\,0, \, \gamma} \ := \ \ell_0^{\,\prime\prime} -2{k}N\bar N +\gamma \ = \
\mathcal{P}_a \mathcal{P}^a-kN\bar N +\gamma \  \approx \ 0 \,
\ee
as a true modified constraint with an arbitrary constant parameter $\gamma$ which will be fixed later.
Let us emphasize once again the important property obtained, the constraint \p{L0-a} has
zero Poisson brackets with $\ell$ and $\tilde \ell$:
\be\lb{F0-ll}
\left\{ \bm{\ell}_{\, 0, \, \gamma},\ell \right\} =\left\{ \bm{\ell}_{\, 0, \, \gamma},\tilde \ell \right\} =0 \,.
\ee
We will show that this quantity can be associated with true constraint $\bm{\ell}_{0}$ at some fixed $\gamma.$

Let us turn to the constraints $\ell\approx0$, $\tilde \ell\approx0$.
Relation \p{PB-ll-T00} shows that their Poisson bracket can not reproduce the
quantity $\bm{\ell}_{\,0, \, \gamma}$ at any $\gamma$. Therefore, if we desire to obtain the true constraints $\bm{\ell}$ and $\tilde{\bm{\ell}}$ from
the quantities $\ell\approx0$ and $\tilde \ell\approx0$, the latter should be modified in such a way
that their Poisson bracket is equal to a linear combination of new constraints, unlike the Poisson bracket \p{PB-ll-T00}.

The desired modification is realized by adding
the terms containing $N\bar N$ to the constraints  $\ell\approx0$, $\tilde \ell\approx0$. As a result, we arrive to the following functions:
\be\lb{LL-a}
\bm{\ell}_{\,\beta_1} \ := \ \ell +\beta_1 N\bar N \  \approx \ 0 \,,\qquad\quad
\tilde{\bm{\ell}}_{\,\beta_2} \ := \ \tilde \ell +\beta_2 N\bar N \  \approx \ 0 \,,
\ee
where $\beta_1$ and  $\beta_2$ are some arbitrary constants.
In the general case, these constants are not equal to each other, $\beta_1\neq\beta_2$, since the constraints \p{LL-a} are real and are not related one to another.

Now we require that the quantities $\bm{\ell}_{\,0, \, \gamma}$, $\bm{\ell}_{\,\beta_1}$, $\tilde{\bm{\ell}}_{\,\beta_2}$ form the closed algebra. We will show that this requirement completely fixes the constants $\gamma$, $\beta_1$, $\beta_2$.

Since $\left\{M_{ab},K \right\} =0$, the constraint \p{L0-a} has
zero Poisson brackets with constraints \p{LL-a}: $\left\{
\bm{\ell}_{\,0,\, \gamma},\bm{\ell}_{\,\beta_1} \right\} = \left\{
\bm{\ell}_{\,0, \, \gamma},\tilde{\bm{\ell}}_{\,\beta_2}  \right\}
=0$. The Poisson bracket of the quantities \p{LL-a} equals
\be\lb{alg-LL-a}
\left\{\bm{\ell}_{\,\beta_1},\tilde{\bm{\ell}}_{\,\beta_2}  \right\}
=-K\bm{\ell}_{\,0}+ K\Big(\beta_2
\bm{\ell}_{\,\beta_1}+\beta_1\tilde{\bm{\ell}}_{\,\beta_2}  \Big)
-2\Big(k+ \beta_1\beta_2  \Big)KN\bar N+
\Big(\gamma+\bm{\mu}\left(\beta_1+\beta_2\right)\Big)K\,.
\ee
To
close off this algebra, the third and fourth terms on the
right-hand side must be equal to zero. This leads to the following
conditions on the constants $\gamma$, $\beta_1$ and  $\beta_2$:
\be\lb{val-b} \beta_1\beta_2={}-k\,,\qquad \gamma={}-\bm{\mu}\left(
\beta_1+\beta_2\right)\,. \ee

Moreover, we can consider the case when $\beta_1=\beta_2$. Then, the conditions \p{val-b} yield
\be\lb{val-ba}
\beta_1=\beta_2=\varkappa\,,\qquad \gamma={}-2\varkappa\bm{\mu}\,.
\ee

As a result, we obtain that the bosonic constraints \p{L0-a}, \p{LL-a} have the form
\bea
\lb{L0-f}
{\bm{\ell}}_{\,0} &=&
\mathcal{P}_a \mathcal{P}^a+\varkappa^2N\bar N -2\varkappa\bm{\mu} \  \approx \ 0 \,
\\ [6pt]
\lb{LL-f} {\bm{\ell}} &=& \left(\xi\sigma^a \bar\xi
\right)\mathcal{P}_a+\varkappa N\bar N-\bm{\mu}  \  \approx \ 0
\,,\\ [6pt] \lb{LtL-f} \tilde{\bm{\ell}} &=&
\left(\bar\pi\tilde\sigma^a \pi \right)\mathcal{P}_a+\varkappa N\bar
N -\bm{\mu}  \  \approx \ 0 \,.
\eea
The fourth bosonic constraint
is the constraint \p{const-sp-3-curv} having the form
\be \lb{U-f}
{\mathcal{U}} \,\  = \,\  N-\bar N \  \approx \ 0 \,.
\qquad\qquad\qquad
\ee
The nonvanishing Poisson bracket of the
constraint \p{L0-f}-\p{U-f} is
\be\lb{alg-LL-f}
\left\{{\bm{\ell}},\tilde{\bm{\ell}} \right\}
={}-K{\bm{\ell}}_{\,0}+ \varkappa K\Big(
{\bm{\ell}}+\tilde{\bm{\ell}} \Big) \,.
\ee

Note that the constraints  \p{L0-f}, \p{LL-f}, \p{LtL-f} in the AdS space can be represented
in the form similar to the corresponding  constraints \p{const-sp}, \p{const-sp-1}, \p{const-sp-2} in flat space.
For this we introduce the vector quantity
\be\lb{def-Pi}
\Pi_a \ := \ \mathcal{P}_a - \frac{\varkappa}{2}\left( \xi\sigma_a\bar\xi + \bar\pi\tilde\sigma_a \pi\right).
\ee
Using vector \p{def-Pi}, the constraints \p{LL-f}, \p{LtL-f} are represented as
\begin{eqnarray}
{\bm{\ell}} &=& \Pi_a \left(\xi\sigma^a \bar\xi \right)-\bm{\mu}  \ \approx \ 0  \,,  \label{constr-sp-1}\\ [5pt]
\tilde{\bm{\ell}} &=& \Pi_a \left(\bar\pi\tilde\sigma^a \pi \right) -\bm{\mu} \ \approx \ 0  \,.  \label{constr-sp-2}
\end{eqnarray}
Moreover, from the explicit form \p{def-Pi} for the vector $\Pi_a$ we have the equality
\be\label{eq-constr-sp}
\Pi_a \Pi^a \ = \ {\bm{\ell}}_{\,0}  - \varkappa\left( {\bm{\ell}}+\tilde{\bm{\ell}}\right)\,.
\ee
That is, the set of constraints  \p{L0-f}, \p{LL-f}, \p{LtL-f} and the set of constraints consisting of both the constraint
\be\label{constr-sp}
\hat{\bm{\ell}}_{\,0} \ = \ \Pi_a \Pi^a \ \approx \ 0
\ee
and the constraints \p{constr-sp-2}, \p{constr-sp-2} are equivalent.

The form of the constraints  \p{constr-sp}-\p{constr-sp-2} in the AdS space  is similar to the form of the constraints
\p{const-sp}-\p{const-sp-2} in flat space.
The constraints  \p{constr-sp}-\p{constr-sp-2} turn
into the constraints  \p{const-sp}-\p{const-sp-2} through the replacement
$p_a \leftrightarrow \Pi_a$.

One can see that the transition from the system with bosonic
constraints \p{const-sp-curv}-\p{const-sp-2-curv} to the system with
bosonic constraints \p{L0-f}-\p{LtL-f} is achieved in the case of
AdS superspace by adding the extra term
\begin{equation}
\label{L-curl-dop}
L_{extra} \ = \ \frac{\varkappa}{2}\,\dot z^M
E_M{}^a \left[(\xi\sigma_a \bar\xi ) + (\bar\pi\tilde\sigma_a \pi)
\right]
\end{equation}
to the Lagrangian \p{L-super-curl}. After the indicated modification
of the three bosonic constraints, we arrive at the following total
Lagrangian:
\begin{equation}
\label{L-super-curl-tot} L_{tot} \ = \ L  \   +  L_{extra}\,.
\end{equation}
This Lagrangian has the same fourth bosonic constraint \p{U-f} and
the same fermionic constraints \p{cov-mom-odd}. As we will see in
the next section, no further modification of the fermionic
constraints is required and the Lagrangian \p{L-super-curl-tot} is
the final Lagrangian of the system under consideration.

\subsection{Structure of the fermionic constraints}

Modified bosonic constraints must have zero Poisson brackets (in the
weak sense) with fermionic quantities \p{cov-mom-odd}:
$\mathscr{D}_{\alpha}$ and $\bar{\mathscr{D}}_{\dot\alpha}$. This
will be a necessary condition for the quantities
$\mathscr{D}_{\alpha}\approx 0$ and
$\bar{\mathscr{D}}_{\dot\alpha}\approx 0$ to be fermionic
constraints that do not require additional modification of bosonic
constraints.

Using \p{PB-PP}-\p{PB-PbD},
we obtain that the Poisson brackets of the constraints \p{L0-f}-\p{LtL-f} with
\p{cov-mom-odd} have the form (here the identities
$\Omega_A^{ab}\mathcal{P}_a\mathcal{P}_b\equiv 0$
are taken into account):
\bea
\left\{ \bm{\ell}_{\,0} \ ,\ \mathscr{D}_{\beta}\right\} &=&{}
-i\,\mathcal{P}^a\left( R_{a\beta}{}^{cd}M_{cd} \ + \ 2T_{a \beta}{}^{C}\mathcal{P}_C\right)
+2\,\mathcal{P}^a\Omega_{a\beta}{}^{\gamma}{\mathscr{D}}_\gamma\,, \lb{L0-Pa} \\
\left\{ \bm{\ell} \ ,\ \mathscr{D}_{\beta} \right\} &=&{}
-i\left(\xi\sigma^a \bar\xi \right)\left(\frac{1}{2}\,R_{a\beta}{}^{cd}M_{cd} \ + \ T_{a\beta}{}^{C}\mathcal{P}_C\right)
+\left(\xi\sigma^a \bar\xi \right)\Omega_{a\beta}{}^{\gamma}{\mathscr{D}}_\gamma\,, \lb{L-Pa}\\
\left\{ \tilde{\bm{\ell}} \ ,\ \mathscr{D}_{\beta} \right\} &=&{}
-i\left(\bar\pi\tilde\sigma^a \pi \right)\left(\frac{1}{2}\,R_{a\beta}{}^{cd}M_{cd} \ + \ T_{a\beta}{}^{C}\mathcal{P}_C\right)
+\left(\bar\pi\tilde\sigma^a \pi \right)\Omega_{a\beta}{}^{\gamma}{\mathscr{D}}_\gamma
\lb{Lt-Pa}
\eea
and c.c. brackets.\footnote{Let us emphasize the presence of the minus sign in the superconnection matrix
(see, e.g., \cite{Ideas}):
$$
\Omega_{AB}{}^C = \left(
\begin{array}{ccc}
\Omega_{A}{}_b{}^c{} & 0 & 0 \\
0 & \Omega_{A}{}_\beta{}^\gamma & 0 \\
0 & 0 & -\Omega_{A}{}^{\dot\beta}{}_{\dot\gamma} \\
\end{array}
\right).
$$} In AdS superspace, the nonvanishing components of the supercurvature and supertension
are given in \p{deS-curv}. Therefore, in this case, the Poisson
brackets \p{L0-Pa}-\p{Lt-Pa} take the form \be\lb{L0-L-Pa}
\begin{array}{rcl}
\left\{ \bm{\ell}_{\,0}\ ,\ {\mathscr{D}}_{\beta} \right\} &=&{} \displaystyle
i \varkappa\mathcal{P}_a\sigma^a_{\beta\dot\gamma}\bar{\mathscr{D}}^{\dot\gamma}+
2\,\mathcal{P}^a\Omega_{a\beta}{}^{\gamma}{\mathscr{D}}_\gamma\,, \\ [7pt]
\left\{ \bm{\ell} \ ,\ {\mathscr{D}}_{\beta} \right\} &=&{}  \displaystyle
-i \varkappa\xi_\beta\bar\xi_{\dot\gamma}\bar{\mathscr{D}}^{\dot\gamma}
+\left(\xi\sigma^a \bar\xi \right)\Omega_{a\beta}{}^{\gamma}{\mathscr{D}}_\gamma\,, \\ [7pt]
\left\{ \tilde{\bm{\ell}} \ ,\ {\mathscr{D}}_{\beta} \right\} &=&{}  \displaystyle
-i \varkappa\pi_\beta\bar\pi_{\dot\gamma}\bar{\mathscr{D}}^{\dot\gamma}
+\left(\bar\pi\tilde\sigma^a \pi \right)\Omega_{a\beta}{}^{\gamma}{\mathscr{D}}_\gamma
\end{array}
\ee and c.c. brackets. Thus, after modification of the bosonic
constraints by adding extra terms with the operators $N$ and $\bar N$,
the new bosonic constraints  \p{L0-f}-\p{LtL-f} have vanishing (in the
weak sense) brackets with the fermionic constraints
$\mathscr{D}_{\alpha}{\approx}\,0$ and
$\bar{\mathscr{D}}_{\dot\alpha}{\approx}\,0$. Besides, the
constraint ${\mathcal{U}}{\approx}\,0$ defined in \p{U-f} has also
zero brackets with $\mathscr{D}_{\alpha}$ and
$\bar{\mathscr{D}}_{\dot\alpha}$: \be \left\{ {\mathcal{U}}\ ,\
\mathscr{D}_{\alpha} \right\} =0\,,\qquad \left\{ {\mathcal{U}}\ ,\
\bar{\mathscr{D}}_{\dot\alpha} \right\} =0\,. \ee

Hence, the fermionic constraints under consideration
$\mathscr{D}_{\alpha}{\approx}\,0$ and
$\bar{\mathscr{D}}_{\dot\alpha}{\approx}\,0$ commute (in the weak sense)
with all bosonic constraints in terms of the Poisson brackets:
\begin{itemize}
\item
all bosonic constraints \p{L0-f}-\p{LtL-f}, \p{U-f}
are of the first class in the supersymmetric system under consideration;
\item
in the flat limit, the constraints $\mathscr{D}_{\alpha}{\approx}\,0$, $\bar{\mathscr{D}}_{\dot\alpha}{\approx}\,0$ \p{cov-mom-odd}
transform into the flat fermionic constraints $D_{\alpha}{\approx}\,0$, $\bar{D}_{\dot\alpha}{\approx}\,0$ \p{D-constr}.
\end{itemize}

Now let us analyze the classes of the fermionic constraints
$\mathscr{D}_{\alpha}{\approx}\,0$ and
$\bar{\mathscr{D}}_{\dot\alpha}{\approx}\,0$. Using
\p{PB-DD}-\p{PB-DbD}, we obtain that the Poisson brackets of
the fermionic constraints $\mathscr{D}_{\alpha}\approx0$ and
$\bar{\mathscr{D}}_{\dot\alpha}\approx0$ in AdS space have the form
(on the surface of the constraints $\mathscr{D}_{\alpha}\approx0$
and $\bar{\mathscr{D}}_{\dot\alpha}\approx0$) \be\lb{DB-f-constr}
\left(
\begin{array}{cc}
\left\{ \mathscr{D}_{\alpha} ,\mathscr{D}^{\beta} \right\} & \left\{ \mathscr{D}_{\alpha} ,\bar{\mathscr{D}}_{\dot\beta} \right\} \\
[7pt]
\left\{ \bar{\mathscr{D}}^{\dot\alpha},\mathscr{D}^{\beta} \right\} & \left\{ \bar{\mathscr{D}}^{\dot\alpha} ,\bar{\mathscr{D}}_{\dot\beta} \right\} \\
\end{array}
\right)={}-2i\, \mathcal{C}_{\underline{A}}{}^{\underline{B}}\,,
\ee
where
\be\lb{def-f-C}
\mathcal{C}_{\underline{A}}{}^{\underline{B}}:=\left(
\begin{array}{cc}
{}-2i \varkappa M_{\alpha}{}^{\beta}  & (\sigma^c)_{\alpha\dot\beta}\mathcal{P}_c \\
[7pt]
(\tilde\sigma^c)^{\dot\alpha\beta}\mathcal{P}_c & 2i \varkappa\bar M^{\dot\alpha}{}_{\dot\beta} \\
\end{array}
\right)\,.
\ee
In \p{DB-f-constr}, \p{def-f-C} the index $\underline{A}$ is the 4-component Dirac-spinor index:
\be
X_{\underline{A}}=
\left(\!
\begin{array}{c}
X_{\alpha} \\
X^{\dot\alpha} \\
\end{array}
\!\right) , \qquad
X^{\underline{A}}=\left(X^{\alpha},X_{\dot\alpha}\right); \ee and
$M_{\alpha\beta}$ and $\bar M_{\dot\alpha\dot\beta}$ are defined in
\p{M-sp-comp}. As a result, the $4\times 4$ matrix equality
\p{DB-f-constr} is represented as
\be\lb{DB-f-constr-m}
\left\{
\mathscr{D}_{\underline{A}} ,\mathscr{D}^{\underline{B}} \right\}\
=\ {}-2i\, \mathcal{C}_{\underline{A}}{}^{\underline{B}}\,.
\ee Note
that after using the equalities (see \p{M-sp-s})
\be
{M}_{\alpha\beta} = \frac12\,M_{ab}(\sigma^{ab})_{\alpha\beta}
\,,\qquad \bar{M}^{\dot\alpha\dot\beta} = {} - \frac12\,
M_{ab}(\tilde{\sigma}^{ab})_{\dot\alpha\dot\beta}\,,
\ee
the matrix
\p{def-f-C} looks like
\begin{equation}
\label{S-P-M}
\mathcal{C}\ = \ \mathcal{P}_a \gamma^a  -
i M_{ab} \Sigma^{ab} \,,
\end{equation}
where
\be
(\gamma^a)_{\underline{A}}{}^{\underline{B}} =
\left(
\begin{array}{cc}
0  & (\sigma^a)_{\alpha\dot\beta} \\
[7pt]
(\tilde\sigma^a)^{\dot\alpha\beta} & 0 \\
\end{array}
\right)
\ee
are the Dirac matrices and
\be
\Sigma^{ab} =
{}-\frac14\, [\gamma^a,\gamma^b]\,,\qquad
( \Sigma^{ab})_{\underline{A}}{}^{\underline{B}} =
\left(
\begin{array}{cc}
{}(\sigma^{ab})_{\alpha}{}^{\beta}  & 0 \\
[7pt]
0 & (\tilde\sigma^{ab})^{\dot\alpha}{}_{\dot\beta} \\
\end{array}
\right)
\ee
are the Lorentz generators in the Dirac-spinor representation.

To calculate the determinant of the matrix \p{DB-f-constr}, we use the relation
\be\lb{example}
\det\left(
\begin{array}{cc}
A_{\alpha}{}^{\beta}  & B_{\alpha\dot\beta}  \\
[7pt]
B^{\dot\alpha\beta} & \bar A^{\dot\alpha}{}_{\dot\beta} \\
\end{array}
\right)={} \frac14\left(B_{\alpha\dot\beta}B^{\dot\beta\alpha} \right)^2
+\frac14\,A^{\alpha\beta}A_{\alpha\beta}\,\bar A_{\dot\alpha\dot\beta}\bar A^{\dot\alpha\dot\beta}
-A_{\alpha}{}^{\beta}B_{\beta\dot\gamma}\bar A^{\dot\gamma}{}_{\dot\delta}B^{\dot\delta\alpha}\,,
\ee
where $A_{\alpha\beta}=\epsilon_{\beta\gamma}A_{\alpha}{}^{\gamma}$,
$A^{\alpha\beta}=\epsilon^{\alpha\gamma}A_{\gamma}{}^{\beta}$,
$\bar A^{\dot\alpha\dot\beta}=\epsilon^{\dot\beta\dot\gamma}\bar A^{\dot\alpha}{}_{\dot\gamma}$,
$\bar A_{\dot\alpha\dot\beta}=\epsilon_{\dot\alpha\dot\gamma}\bar A^{\dot\gamma}{}_{\dot\beta}$,
$B^{\dot\alpha\beta}=\epsilon^{\dot\alpha\dot\gamma}\epsilon^{\beta\delta} B_{\delta\dot\gamma}$
and all matrix components are $c$-numbers.
Using \p{example} and the identity
$$
(\xi\sigma^a\bar\pi)\mathcal{P}_{a}(\pi\sigma^b\bar\xi)\mathcal{P}_{b}=
(\xi\sigma^a\bar\xi)\mathcal{P}_{a}(\pi\sigma^b\bar\pi)\mathcal{P}_{b}+N\bar N\mathcal{P}^{a}\mathcal{P}_{a}\,,
$$
we obtain that
the determinant of the Poisson brackets matrix \p{DB-f-constr} has the form
\be\lb{det-f-c}
\det\mathcal{C} \ = \ \left(\mathcal{P}^{a}\mathcal{P}_{a}\right)^2 \ + \
|\varkappa|^4 N^2\bar N^2  \ - \ 4|\varkappa|^2(\xi\sigma^a\bar\xi)\mathcal{P}_{a}(\pi\sigma^b\bar\pi)\mathcal{P}_{b}
 \ - \ 2|\varkappa|^2N\bar N\,\mathcal{P}^{a}\mathcal{P}_{a}
\,.
\ee

Using the expressions for $\mathcal{P}^{a}\mathcal{P}_{a}$,
$(\xi\sigma^a\bar\xi)\mathcal{P}_{a}(\pi\sigma^b\bar\pi)\mathcal{P}_{b}$
from \p{L0-f}-\p{LtL-f},
one obtains
\be\lb{det-f-1}
\det\mathcal{C}\ =\ (\bm{\ell}_{\,0})^2
-4\varkappa\left(\varkappa N\bar N - \bm{\mu}\right)\left[\bm{\ell}_{\,0}-\varkappa\left(\bm{\ell} +\tilde{\bm{\ell}}\right)\right]
-4\varkappa^2 \bm{\ell} \tilde{\bm{\ell}} \,.
\ee
Therefore, on the surface of bosonic constraints,
the determinant of the matrix constructed from the Poisson brackets of fermionic  constraints is equal to zero:
\be\lb{det-f-zero}
\det\mathcal{C}\ \approx\ 0\,.
\ee
Thus, in the case under consideration, four constraints
$\mathscr{D}_{\alpha}\approx0$ and $\bar{\mathscr{D}}_{\dot\alpha}\approx0$
form a mixture of first- and second-class constraints.
Moreover, of these four fermionic constraints, at least one constraint is the first-class one.

\subsection{Division of the fermionic constraints into first- and second-class constraints}

In this subsection we will consider the division of the four fermionic constraints
$\mathscr{D}_{\alpha}\approx0$ and $\bar{\mathscr{D}}_{\dot\alpha}\approx0$
into the first- and second-class constraints. First of all, let us find the numbers of separate first- and
separate second-class constraints that are present in the above fermionic constraints.

Considering the matrix $\mathcal{C}$ \p{S-P-M}, one can see that
the matrix $\gamma^0\mathcal{C}$ is Hermitian and therefore is diagonalized by a unitary transformation.
The corresponding characteristic equation that determines the eigenvalues and rank of the matrix
$\gamma^0\mathcal{C}$ has the form
\be\lb{det-f-lambda-g}
\det\left(\gamma^0\mathcal{C} -\lambda \mathbf{1}_4\right)\ =\ 0\,,
\ee
where the matrix $\mathbf{1}_4$
is the identity $4\times 4$ matrix and the number $\lambda$ is the eigenvalue of the matrix $\mathcal{C}$.
Equation \p{det-f-lambda-g} is equivalent to the equation
\be\lb{det-f-lambda}
\det\left(\mathcal{C} -\lambda \gamma^0\right)\ =\ 0\,.
\ee

Equation \p{det-f-lambda} is obtained from equation \p{det-f-zero} by substituting
\be
\mathcal{P}_{0} \ \ \to \ \ \mathcal{P}_{0}-\lambda \,.
\ee
Therefore, one gets $\bm{\ell}_{\,0}\to \bm{\ell}_{\,0}+2\mathcal{P}_{0}\lambda-\lambda^2$,
$\bm{\ell}\to \bm{\ell}+(\xi\sigma_0\bar\xi)\lambda$,
$\tilde{\bm{\ell}}\to \tilde{\bm{\ell}}+(\pi\sigma_0\bar\pi)\lambda$.
Then, using \p{det-f-1}, we obtain
\be\lb{det-C-lambda-0}
\det\left(\mathcal{C} -\lambda \gamma^0\right)\ = \
\det \mathcal{C}\ +\ a\lambda\ +\ b\lambda^2\ -\ 4 \mathcal{P}_{0} \lambda^3\ +\ \lambda^4\,,
\ee
where
\bea
a&=& 4 \mathcal{P}_{0}\bm{\ell}_{\,0} \ -\  4\varkappa^2(\pi\sigma_0\bar\pi)\bm{\ell}
\ -\  4\varkappa^2(\xi\sigma_0\bar\xi)\tilde{\bm{\ell}}
\ +\ 8\varkappa\left(\bm{\mu}-\varkappa N\bar N \right)\Pi_0 \,, \label{a-def}\\ [7pt]
b&=& 4(\mathcal{P}_{0})^2\ + \ 2\bm{\ell}_{\,0} \ + \ \varkappa\left(\varkappa N\bar N - \bm{\mu}\right)
\ - \ (\xi\sigma_0\bar\xi)(\pi\sigma_0\bar\pi)\,,
\eea
where $\Pi_a$ is defined in \p{def-Pi}.

On the constraint surface the quantity \p{a-def} takes the form $a= 8\varkappa\left(\bm{\mu}-\varkappa N\bar N\right)\Pi_0$.
Further we assume that $\left(\bm{\mu}-\varkappa N\bar N\right)\neq 0$.
Otherwise we would have one more bosonic constraint $N\bar N-\bm{\mu}/\varkappa\approx0$,
which, together with the existing constraint \p{U-f}, would completely fix the values of $N$ and $\bar N$, which
corresponds to a particle of finite spin (helicity) (see, e.g., the consideration of such systems in
\cite{BFIK-2024}).
Also note that on the constraint surface \p{constr-sp} the vector $\Pi_a$ is light-like.
Therefore,  $\Pi_0\neq 0$. As result, $a\neq 0$.
Therefore, the Hermitian matrix $\gamma^0\mathcal{C}$ has only one zero eigenvalue.
Thus, we have one first-class constraint and three second-class constraints
among four fermionic constraints
$\mathscr{D}_{\alpha}\approx0$ and $\bar{\mathscr{D}}_{\dot\alpha}\approx0$.

We now proceed to explicitly divide the above four fermionic constraints into
one first-class constraint and three second-class ones. The division procedure is in principle  similar
to what was done in the flat case (see  \p{sG-constr}, \p{fF-constr}).
But now we use the identities
\be
\label{delta-curv}
\delta_{\alpha}^{\beta}=\frac{1}{\bm{\mu}}\left[\xi_{\alpha}
(\bar\xi\tilde \Pi)^{\beta} +(\Pi\bar\xi)_{\alpha}\xi^{\beta}\right],\qquad
\delta_{\dot\alpha}^{\dot\beta}=\frac{1}{\bm{\mu}}\left[\bar\xi_{\dot\alpha}(\tilde \Pi\xi)^{\dot\beta}
+(\xi \Pi)_{\dot\alpha}\bar\xi^{\dot\beta}\right],
\ee
which are a curve superspace generalization of flat identities \p{delta} and are fulfilled on the constraint
surface \eqref{constr-sp-1} and \eqref{constr-sp}.

Taking into account identities (\ref{delta-curv}), we consider two constraints
\begin{equation}\lb{sG-constr-c}
\mathcal{G}:=\xi^{\alpha}{\mathscr{D}}_{\,\alpha}\approx 0\,,
\qquad   \bar{\mathcal{G}}:= \bar{\mathscr{D}}_{\,\dot\alpha}\bar\xi^{\dot\alpha}\approx 0\,,
\qquad  (\mathcal{G})^* ={}-\bar{\mathcal{G}}
\end{equation}
and two constraints
\begin{equation}\lb{fF-constr-c}
\mathcal{F}:=\Pi^b\,\bar\xi_{\dot\alpha}\sigma_b^{\dot\alpha\alpha}{\mathscr{D}}_{\,\alpha}\approx 0\,,
\qquad   \bar{\mathcal{F}}:= \bar{\mathscr{D}}_{\,\dot\alpha}\sigma_b^{\dot\alpha\alpha}\xi_{\alpha}\,\Pi^b\approx 0\,,
\qquad  (\mathcal{F})^* ={}-\bar{\mathcal{F}}\,.
\end{equation}
The nonvanishing Poisson brackets of the constraints \p{sG-constr-c}, \p{fF-constr-c} in the superspace with
the supercurvature and supertorsion \p{deS-curv} have the form:
\be
\left\{ \mathcal{G}\ ,\ \bar{\mathcal{G}} \right\}  \ =\ {}
-2i\left(\xi\mathcal{P}\bar\xi\right) ,\qquad
\left\{ \mathcal{G}\ ,\ {\mathcal{G}} \right\}  \ =\ {}0\,,\qquad
\left\{ \bar{\mathcal{G}}\ ,\ \bar{\mathcal{G}} \right\}  \ =\ {}0\,,\qquad
\ee
\bea
&& \left\{ \mathcal{G}\ ,\ {\mathcal{F}} \right\}
\ ={}\  2\varkappa N\left(\xi\Pi\bar\xi\right)\ +\  \varkappa\, \mathcal{G}\bar{\mathcal{G}}
\ -\  \varkappa\bar N {\mathscr{D}}^{\,\alpha}{\mathscr{D}}_{\,\alpha}\,,
\\ [5pt]
&& \left\{ \bar{\mathcal{G}}\ ,\ \bar{\mathcal{F}} \right\}
\ ={}\  -2\varkappa \bar N\left(\xi\Pi\bar\xi\right)\ -\  \varkappa\, \mathcal{G}\bar{\mathcal{G}}
\ +\  \varkappa N \bar{\mathscr{D}}_{\,\dot\alpha}\bar{\mathscr{D}}^{\,\dot\alpha}\,,
\\ [5pt]
&& \left\{ \mathcal{G}\ ,\ \bar{\mathcal{F}} \right\}
\ ={}\  2i\varkappa N\,\big(\xi\Pi\bar\pi \big)\ +\  \varkappa\, \mathcal{G}\bar{\mathcal{G}}\,,
\\ [5pt]
&& \left\{ \bar{\mathcal{G}}\ ,\ {\mathcal{F}} \right\}
\ ={}\  2i\varkappa \bar N\,\big(\pi\Pi\bar\xi \big)\ +\  \varkappa\, \mathcal{G}\bar{\mathcal{G}}
\,,
\eea
\bea
&& \left\{ \mathcal{F}\ ,\ {\mathcal{F}} \right\} \  =\
{}-2\varkappa\, \bar{\mathcal{G}}\mathcal{F} \ +\  2\varkappa\, (\pi^{\alpha}{\mathscr{D}}_{\,\alpha})\mathcal{F}
\ +\ 4\varkappa \left(\pi\Pi\bar\xi\right)\left(\xi\Pi\bar\xi\right)\,,
\\ [5pt]
&& \left\{ \bar{\mathcal{F}}\ ,\ \bar{\mathcal{F}} \right\} \  =\
{}-2\varkappa\, {\mathcal{G}}\bar{\mathcal{F}} \ +\  2\varkappa\, (\bar{\mathscr{D}}_{\,\dot\alpha}\bar\pi^{\dot\alpha})\bar{\mathcal{F}}
\ -\ 4\varkappa \left(\xi\Pi\bar\pi\right)\left(\xi\Pi\bar\xi\right)\,,
\\ [5pt]
&& \left\{ \mathcal{F}\ ,\ \bar{\mathcal{F}} \right\} \  =\ {}
-2i\big(\xi\Pi\tilde{\mathcal{P}}\Pi\bar\xi\big) \ + \
\varkappa\big({\mathscr{D}}^{\,\alpha}{\mathscr{D}}_{\,\alpha} - \bar{\mathscr{D}}_{\,\dot\alpha}\bar{\mathscr{D}}^{\,\dot\alpha}\big)\big(\xi\Pi\bar\xi\big)
\ + \ \varkappa\big({\mathscr{D}}\Pi\bar{\mathscr{D}}\big) u \,.
\eea

Taking into account the equality $\big(\xi\Pi\tilde{\mathcal{P}}\Pi\bar\xi\big)=\big(\xi\mathcal{P}\bar\xi\big)\Pi^a\Pi_a-
2\big(\xi\Pi\bar\xi\big)\Pi^a\mathcal{P}_a \approx{} -2\varkappa\mu^2$ and also \p{constr-sp-m-a},
we obtain that on the constraint surface the Poisson brackets
of the fermionic constraints have the form
\be\lb{PB-GbG-c}
\big\{ \mathcal{G} ,   \bar{\mathcal{G}} \big\}   \approx   {}
-2i\left(\bm{\mu}-\varkappa N\bar N\right),\qquad
\big\{ \mathcal{G} ,   {\mathcal{G}} \big\}   \approx0\,,\qquad
\big\{ \bar{\mathcal{G}} ,   \bar{\mathcal{G}} \big\}   \approx0\,,
\ee
\be\lb{PB-GbG-F-c}
\begin{array}{c}
\big\{ \mathcal{G}  ,  {\mathcal{F}} \big\}
\approx {}  2\varkappa \mu N\,,
\qquad \big\{ \mathcal{G} ,  \bar{\mathcal{F}} \big\}
\approx {} 2ic\varkappa\mu N
\,,
\\ [6pt]
\big\{ \bar{\mathcal{G}} ,  {\mathcal{F}} \big\}
\approx {}   2i\bar c\varkappa \mu\bar N \,,
\qquad
\big\{ \bar{\mathcal{G}} ,  \bar{\mathcal{F}} \big\}
\approx {} - 2\varkappa \mu \bar N \,,
\end{array}
\ee
\be\lb{PB-FbF-c}
\big\{ \mathcal{F} , {\mathcal{F}} \big\}   \approx
{}4\bar c\varkappa \mu^2\,,
\qquad \big\{ \bar{\mathcal{F}} , \bar{\mathcal{F}} \big\} \approx
{}-4 c\varkappa \mu^2\,,
\qquad \big\{ \mathcal{F} , \bar{\mathcal{F}} \big\}   \approx  {}
4i\varkappa \mu^2 \,.
\ee

{}Using relations \p{PB-GbG-F-c} and \p{PB-FbF-c}, we obtain that the constraint
\be\lb{def-F1}
\mathcal{F}_1 := \mathcal{F}  + \bar c i \bar{\mathcal{F}}
\ee
has the vanishing Poisson brackets with all fermionic constraints
\be
\big\{ \mathcal{G} , {\mathcal{F}}_1 \big\}\approx\big\{ \bar{\mathcal{G}} , {\mathcal{F}}_1 \big\}\approx
\big\{ \mathcal{F} , {\mathcal{F}}_1 \big\}\approx\big\{ \bar{\mathcal{F}} , {\mathcal{F}}_1 \big\}\approx0
\ee
and with itself
(since this constraint is Grassmann odd, this is a non-trivial condition)
\be
\big\{ {\mathcal{F}}_1 , {\mathcal{F}}_1 \big\}\approx0\,.
\ee
Therefore, expression \p{def-F1} is the first-class fermionic constraint.
The realness condition for this constraint looks like:
$(\mathcal{F}_1)^* ={}ci{\mathcal{F}_1}$. When choosing $c=-i$,
the constraint \p{def-F1} is real in conventional form: $(\mathcal{F}_1)^* ={}{\mathcal{F}_1}$.

The remaining constraints, namely, the constraint
\be\lb{def-F2}
\mathcal{F}_2 := \mathcal{F}  -  \bar c i \bar{\mathcal{F}}
\ee
and the constraints $\mathcal{G}$ and $\bar{\mathcal{G}}$, are the second-class constraints.

To obtain the explicit forms of the expressions $\mathcal{G}$ and $\bar{\mathcal{G}}$, we
divide these constraints into real and imaginary parts as follows:
\be\lb{def-G12}
\mathcal{G}_1 := \mathcal{G}  +  c i\, \bar{\mathcal{G}}\,,\qquad
\mathcal{G}_2 := \mathcal{G}  -  c i\, \bar{\mathcal{G}}\,.
\ee
The Poisson brackets of the constraints \p{def-G12} and \p{def-F2} have the form
\be
\big\{ {\mathcal{G}}_1 , {\mathcal{G}}_1 \big\}\approx 4c\left(\bm{\mu}-\varkappa N\bar N\right)\,,\quad
\big\{ {\mathcal{G}}_2 , {\mathcal{G}}_2 \big\}\approx{}- 4c\left(\bm{\mu}-\varkappa N\bar N\right)\,,\quad
\big\{ {\mathcal{G}}_1 , {\mathcal{G}}_2 \big\}\approx 0\,,
\ee
\be
\big\{ {\mathcal{F}}_2 , {\mathcal{F}}_2 \big\}\approx 16\bar c \varkappa \bm{\mu}^2\,,\quad
\big\{ {\mathcal{G}}_1 , {\mathcal{F}}_2 \big\}\approx{}0\,,\quad
\big\{ {\mathcal{G}}_2 , {\mathcal{F}}_2 \big\}\approx 4 \varkappa \bm{\mu}(N+\bar N)\,.
\ee

The constraint ${\mathcal{G}}_1$ has splitted off and we clearly see that it is of the second-class.

For the remaining two  constraints ${\mathcal{G}}_2$ and ${\mathcal{F}}_2$,
the Poisson bracket matrix
\be\lb{PB-F2G2}
\left(
\begin{array}{cc}
\left\{ {\mathcal{G}}_2 ,{\mathcal{G}}_2 \right\} & \left\{ {\mathcal{G}}_2 ,{\mathcal{F}}_2 \right\} \\
[7pt]
\left\{ {\mathcal{F}}_2,{\mathcal{G}}_2 \right\} & \left\{ {\mathcal{F}}_2 ,{\mathcal{F}}_2 \right\} \\
\end{array}
\right)\approx{}4
\left(
\begin{array}{cc}
- c\left(\bm{\mu}-\varkappa N\bar N\right) & \varkappa \bm{\mu}(N+\bar N) \\
[7pt]
\varkappa \bm{\mu}(N+\bar N) & 4\bar c \varkappa \bm{\mu}^2 \\
\end{array}
\right)
\ee
has the determinant that is proportional to $\varkappa \bm{\mu}^3$ on the constraints surface and is non-zero.
This shows that the constraints ${\mathcal{G}}_2$ and ${\mathcal{F}}_2$ are  the second-class ones.

As a result, the model of a continuous spin superparticle in $AdS_4$ superspace we have
constructed is described by four bosonic first-class constraints
${\bm{\ell}}_{\,0}\approx0$ \p{L0-f}, ${\bm{\ell}}\approx0$ \p{LL-f},
$\tilde{\bm{\ell}}\approx0$ \p{LtL-f}, ${\mathcal{U}}\approx0$ \p{U-f}
and four fermionic  constraints
$\mathscr{D}_{\alpha}\approx0$, $\bar{\mathscr{D}}_{\dot\alpha}\approx0$ \p{cov-mom-odd}.
The latter are a mixture of one first-class constraint $\mathcal{F}_1\approx0$ \p{def-F1}
and three second-class constraints $\mathcal{F}_2\approx0$ \p{def-F2},
$\mathcal{G}_1\approx0$, $\mathcal{G}_2\approx0$ \p{def-G12}.

Let us note the difference between the model of a continuous spin superparticle
in flat superspace and in anti-de Sitter superspace. Despite the same type of bosonic constraints
(there are four of them and they are all of the first class),
the fermionic constraints have a different structure.
In the flat case, four fermionic constraints
$D_{\,\alpha}\approx0$, $\bar D_{\,\dot\alpha}\approx0$ \p{D-constr}
form a mixture of two irreducible second-class constraints  $G\approx0$, $\bar G\approx0$ \p{sG-constr} and
two irreducible first-class constraints  $F\approx0$, $\bar F\approx0$ \p{fF-constr}.
In $AdS_4$ superspace, the analogue of the flat constraints
$D_{\,\alpha}\approx0$, $\bar D_{\,\dot\alpha}\approx0$ \p{D-constr} is
the constraints
$\mathscr{D}_{\alpha}\approx0$, $\bar{\mathscr{D}}_{\dot\alpha}\approx0$ \p{cov-mom-odd}.
But now only one of these constraints,
namely, the real part  $\mathcal{F}_1\approx0$ \p{def-F1}, is the first-class constraint.
The imaginary part $\mathcal{F}_2\approx0$ \p{def-F2} and
$\mathcal{G}_1\approx0$, $\mathcal{G}_2\approx0$ \p{def-G12} are the second-class constraints.

\subsection{Fermionic $\kappa$-symmetry in curved superspace}

As is well known, any first-class constraint is a generator of gauge transformations in phase space
(see, e.g., \cite{GT}).
In the case under consideration, we have only one first-class fermionic constraint (\ref{def-F1}).
This constraint generates a local fermionic transformation
of the curved  phase superspace coordinates in the form
\be\lb{kappa-g}
\delta z^M=\kappa\big\{ {\mathcal{F}}_1, z^M \big\}\,,\qquad
\delta \xi^\alpha=\kappa\big\{ {\mathcal{F}}_1, \xi^\alpha \big\}\,,\qquad
\delta \bar\xi^{\dot\alpha}=\kappa\big\{ {\mathcal{F}}_1, \bar\xi^{\dot\alpha} \big\}\,,\qquad
\ee
with $\kappa$ being the only one (real in the case $c=-i$) anticommuting local transformation parameter.
Similar transformations will be held for the momenta $p_{M},\,\pi_{\alpha},\,\bar{\pi}_{\dot{\alpha}}$.
Direct calculations with the use of the explicit expression for the first-class constraint (\ref{def-F1}) lead to the following $\kappa$-symmetry
transformations for the continuous superparticle model in curved superspace:
\bea
\delta z^M&=&{}-\kappa \left[\bar\xi_{\dot\alpha}\tilde\Pi^{\dot\alpha\alpha}E_{\alpha}{}^{M}
+ \bar c i \xi^\alpha\Pi_{\alpha\dot\alpha}E^{\dot\alpha M}\right]
-\kappa \left[(\bar\xi \tilde\sigma^{a}\mathscr{D})
+ \bar c i (\xi\sigma^{a}\bar{\mathscr{D}})\right]E_a{}^{M}\,,
\\ [6pt]
\delta \xi^\alpha&=&{}i\kappa \left[(\bar\xi\tilde\Pi)^{\gamma}\Omega_{\gamma}{}^{\alpha\beta}
+ \bar c i (\xi\Pi)_{\dot\gamma}\Omega^{\dot\gamma\alpha\beta}\right]\xi_\beta
+\kappa \left[(\bar\xi \tilde\sigma^{a}\mathscr{D})
+ \bar c i (\xi\sigma^{a}\bar{\mathscr{D}})\right]\Omega_a{}^{\alpha\beta}\xi_\beta\,,
\\ [6pt]
\delta \bar\xi^{\dot\alpha}&=&{}i\kappa \left[(\bar\xi\tilde\Pi)^{\gamma}\Omega_{\gamma}{}^{\dot\alpha\dot\beta}
+ \bar c i (\xi\Pi)_{\dot\gamma}\Omega^{\dot\gamma\dot\alpha\dot\beta}\right]\bar\xi_{\dot\beta}
+\kappa \left[(\bar\xi \tilde\sigma^{a}\mathscr{D})
+ \bar c i (\xi\sigma^{a}\bar{\mathscr{D}})\right]\Omega_a{}^{\dot\alpha\dot\beta}\bar\xi_{\dot\beta}\,,
\\ [6pt]
\delta \pi_\alpha&=&{}-i\kappa \left[(\bar\xi\tilde\Pi)^{\gamma}\Omega_{\gamma\alpha\beta}
+ \bar c i (\xi\Pi)_{\dot\gamma}\Omega^{\dot\gamma}{}_{\alpha\beta}\right]\pi^\beta
-\kappa \left[(\bar\xi \tilde\sigma^{a}\mathscr{D})
+ \bar c i (\xi\sigma^{a}\bar{\mathscr{D}})\right]\Omega_{a\alpha\beta}\pi^\beta
\\ [5pt]
&&{}+ \kappa\bar c i (\Pi\bar{\mathscr{D}})_{\alpha} \nonumber\,,
\\ [6pt]
\delta \bar\pi_{\dot\alpha}&=&{}i\kappa \left[(\bar\xi\tilde\Pi)^{\gamma}\Omega_{\gamma\dot\alpha\dot\beta}
+ \bar c i (\xi\Pi)_{\dot\gamma}\Omega^{\dot\gamma}{}_{\dot\alpha\dot\beta}\right]\bar\pi^{\dot\beta}
+\kappa \left[(\bar\xi \tilde\sigma^{a}\mathscr{D})
+ \bar c i (\xi\sigma^{a}\bar{\mathscr{D}})\right]\Omega_{a\dot\alpha\dot\beta}\bar\pi^{\dot\beta}
\\ [5pt]
&&{}+ \kappa ({\mathscr{D}}\Pi)_{\dot\alpha} \nonumber\,.
\eea
Note that all terms on the right-hand sides of these transformations, except for the first terms on
each line, contain constraints and are equal to zero in the weak sense. Also, one pays attention that
the additional variables $\xi^\alpha$, $\pi_\alpha$ and c.c. also transform under
$\kappa$-transformations in the model under consideration.

\setcounter{equation}{0}

\section{Summary}

In this paper, we have constructed a new superparticle model that
describes the dynamics of a $4D$, $\mathcal{N}{=}\,1$ continuous spin
superparticle in $AdS_4$ superspace.

The constructed model is a generalization of the continuous-spin superparticle model  in flat $4D$, $\mathcal{N}{=}\,1$
superspace proposed in our paper \cite{BF-25}. We have
started with considering a continuous spin superparticle in an
arbitrary curved $4D$, $\mathcal{N}{=}\,1$ superspace. The analysis showed
that in this case we should restrict the bosonic sector
of the $4D$, $\mathcal{N}{=}\,1$ superspace to the constant curvature
space. As a result of local supersymmetry, the continuous
spin superparticle model turns out to be consistent only in the
background superspace $AdS_4$, whose supercurvature and supertorsion
are determined by expressions \p{deS-curv}. The Lagrangian and the bosonic and fermionic
constraints of the model are constructed in explicit form.

After fixing the type of superspace, the Poisson brackets of the
fermionic constraints $\mathscr{D}_{\alpha}\approx0$,
$\bar{\mathscr{D}}_{\dot\alpha}\approx0$ \p{cov-mom-odd} have the form
\p{DB-f-constr}. A straightforward calculation shows that the rank
of the matrix of these Poisson brackets is
three. Therefore, in contrast to the flat case, only one fermionic constraint
(namely, $\mathcal{F}_1\approx0$ \p{def-F1}) of the four
fermionic constraints is the first-class constraint, while the
other three ($\mathcal{F}_2\approx0$ \p{def-F2} and
$\mathcal{G}_1\approx0$, $\mathcal{G}_2\approx0$ \p{def-G12}) are
the second-class constraints. Let us pay attention to that it is
precisely the presence of commuting spinor additional variables in
the model under consideration that made it possible to carry out
such a separation of four fermionic constraints in a covariant manner.

The presence of unequal numbers of first- and second-class
fermionic constraints may cause some confusion. For example, in the
well-known Brink-Schwarz superparticle model \cite{BSch} the number of
fermionic $\kappa$-symmetries is equal to half the number of
fermionic covariant derivatives (or supersymmetry generators). The
same is in the continuous spin superparticle model in flat
superspace \cite{BF-25}. However, the supersymmetric models with a
different number of fermionic $\kappa$-symmetries were also
considered earlier in the literature, mainly in the context of BPS states (see, e.g.,
\cite{GGHT-2001}). In particular, in
\cite{BL-1999,BLS-2000}, the superparticle models with three
$\kappa$-symmetries were constructed for $4D$, $\mathcal{N}{=}\,1$ case (3/4 BPS states).
The $4D$, $\mathcal{N}{=}\,1$ supersymmetric models with
one $\kappa$-symmetry, which correspond to 1/4 BPS states, were studied
in \cite{DIK-2000,FedZim-2000}. We also note the paper
\cite{BGIK-2002}, where a superparticle model corresponding to the case
of 1/4 BPS states and describing the model with
$1D$, $\mathcal{N}{=}\,8$ supersymmetry was considered.

The model of the superparticle obtained here is very non-trivial for
its quantization. In the flat case \cite{BF-25}, the
number of fermionic constraints was even and it was possible to
separate them into two halves, complex conjugate to each other, and
to carry out quantization according to the Gupta-Bleuler procedure. Now,
an odd number of fermionic second-class constraints requires either the
introduction of Dirac brackets for them (or for some of them) or
the use of a procedure for converting second-class constraints into
first-class ones by introducing additional (odd in our case)
variables. Both of these methods require additional analysis, which
is planned to be considered in subsequent works.

We expect that the specific structure of the fermionic constraints obtained here should lead, after quantization,
to specific superfield equations of motion in $AdS$ superspace.
We assume that the form of such equations is related to specific representations of
the $4D$, $\mathcal{N}{=}\,1$ anti-de Sitter $OSp(1|4)$ supergroup. This assumption certainly deserves
study.

Another extremely interesting and promising direction for generalizing the model considered here 
is the construction of a model of the $4D$, $\mathcal{N}{=}\,2$ continuous spin superparticle. 
At present, almost all issues concerning such a model remain open. 
First, a description of continuous spin representations of the $4D$, $\mathcal{N}{=}\,2$ Poincar\'{e} supergroup 
in terms of the $\mathcal{N}{=}\,2$ superspace has not yet been given.
Second, the procedure for the continuous spin superparticle coupling to the $\mathcal{N}{=}\,2$ $AdS$ superspace 
is not entirely obvious.\footnote{The $4D$, $\mathcal{N}{=}\,2$ $AdS$ superspace has been widely discussed in the literature
(see, e.g., \cite{KTM-2021,KKR-2023,KT-2023,KKR,KR} and  references therein).} 
We believe that the most natural way to construct the $4D$, $\mathcal{N}{=}\,2$ superparticle model can be realized 
in terms of harmonic superspace \cite{GIKOS-1985,GKS-1987,GIOS-1987,GIOS-2001}. 
Related to the problem under consideration here, $\mathcal{N}=2$ higher spin field theories were studied in the recent papers  
\cite{BIZ-2021,BIZ-2022,BIZ-2023,BIZ-2024,BIZ-2025}. 
Of particular relevance to our problems is a recent paper \cite{IZ-2025}, 
which deals with superfield theories with $\mathcal{N}{=}\,2$ $AdS$ supergroup $\mathrm{OSp}(2|4)$  symmetry in harmonic superspace.
These results, in principle can serve as a basis for constructing the $4D$, $\mathcal{N}{=}\,2$ 
continuous spin superparticle model in the $AdS$ harmonic superspace.

\section*{Acknowledgments}
We are grateful to Evgeny Ivanov for useful comments.

\section*{Appendix\,A: \ Vectors $\Pi^a$ and $\mathcal{P}^a$ on the constraints surface }
\def\theequation{A.\arabic{equation}}
\setcounter{equation}0

We introduce the vectors
\be\lb{basis-4}
n_+^a:=\xi\sigma^a\bar\xi \,,\qquad n_-^a:=\pi\sigma^a\bar\pi\,, \qquad
m^a:=\xi\sigma^a\bar\pi \,,\qquad \bar m^a:=\pi\sigma^a\bar\xi\,,
\ee
that form a basis in the vector space. Their nonzero products have the form
\be
(n_+ \!\cdot\! n_-) ={}-2N\bar N\,, \qquad (m \!\cdot\! \bar m) ={}2N\bar N\,,
\ee
where we use the following conventions for the products of 4-vectors: $(n_\pm \!\cdot\! m):=n_\pm ^a m_a$,
etc.

The constraints \p{constr-sp-1} and \p{constr-sp-2} rewritten in the form
(here we consider all relations on the constraint surface; therefore,
we use the sign ``$=$'' instead of the  sign ``$\approx$'' of the weak equality)
\be\lb{constr-sp-12-a}
(\Pi \!\cdot\! n_\pm) = \mu\,,
\ee
where the constraint \p{constr-sp} has the form
\be\lb{constr-sp-a}
(\Pi \!\cdot\! \Pi) = 0 \,.
\ee
with the vector $\Pi_a$ defined in \p{def-Pi}.

The vector $\Pi_a$ \p{def-Pi} can be expanded on the space of the constraints
in the basis vectors \p{basis-4}:
\be\lb{basis-4a}
\Pi^a=A^+n_+^a+A^- n_-^a+Bm^a+\bar B\bar m^a\,,
\ee
where $A^\pm$, $B$ and $\bar B$ are some quantities. Let us define them.

The conditions \p{constr-sp-12-a} lead to $\displaystyle A^\pm={}-\frac{\mu}{2N\bar N}$.
Then, the condition \p{constr-sp-a} gives $\displaystyle B={}\frac{\bar c\mu}{2N\bar N}$ and
$\displaystyle \bar B={}\frac{c\mu}{2N\bar N}$, where $c\bar c=1$.
Thus, on the constraint surface the following decomposition takes place:
\be\lb{basis-4-Pi}
\Pi^a={}-\frac{\mu}{2N\bar N}\big(n_+^a+n_-^a\big)+ \frac{\mu}{2N\bar N}\big(\bar cm^a+c\bar m^a\big)\,.
\ee
It is important that, as follows from \p{basis-4-Pi}, the equalities
$(\Pi \!\cdot\! m) = c\mu$ and $(\Pi \!\cdot\! \bar m) = \bar c\mu$ are satisfied on the constraint surface.
That is, the following constraints:
\be\lb{constr-sp-m-a}
(\xi\Pi\bar\pi) =c\mu\,,\qquad (\pi\Pi\bar\xi) = \bar c\mu
\ee
take place.

Note that the constraint \p{U-f} of the form
$\mathcal{U} = \xi^{\alpha}\pi _{\alpha} - \bar\pi _{\dot\alpha}\bar\xi^{\dot\alpha}\approx \ 0 $
generates the transformations
\be\lb{u1-trans}
\xi^{\alpha}\ \to\ e^{i\phi}\xi^{\alpha}\,,\qquad \bar\xi^{\dot\alpha}\ \to\ e^{-i\phi}\bar\xi^{\dot\alpha}\,,\qquad
\pi _{\alpha}\ \to \ e^{-i\phi}\pi _{\alpha} \,,\qquad \bar\pi _{\dot\alpha}\ \to\ e^{i\phi}\bar\pi _{\dot\alpha}\,,
\ee
where $0\leq\phi<2\pi$ is the local angle parameter.
As we can see from the expressions \p{constr-sp-m-a}, the local $\mathrm{U}(1)$-symmetry
\p{u1-trans} allows us to fix the constant $c$, for example, by taking $c=-i$.

Also note that after using \p{def-Pi} and \p{basis-4-Pi}, we get a useful decomposition
of the covariant momentum in the form
\be\lb{basis-4-calP}
\mathcal{P}^a={}\frac12\left(\varkappa-\frac{\mu}{2N\bar N}\right)\big(n_+^a+n_-^a\big)+
\frac{\mu}{2N\bar N}\big(\bar cm^a+c\bar m^a\big).
\ee

\section*{Appendix\,B: \ Poisson brackets of covariant momenta \\
\phantom{.} \qquad \qquad \quad\ \  in AdS superspace  }
\def\theequation{B.\arabic{equation}}
\setcounter{equation}0

\quad\,
In this Appendix we present the Poisson brackets in AdS superspace
for basic quantities.

Using the direct calculation and relations
$$
\left\{ M_{ab}\ ,\ (\xi\sigma^c\bar\xi) \right\} \ = \
\delta_a^c(\xi\sigma_b\bar\xi)\ -\ \delta_b^c(\xi\sigma_a\bar\xi)\,,
$$
we obtain that
in AdS superspace the Poisson brackets \p{cal-Pa-alg}  take the form:
\be
\lb{PB-PP}
\left\{ \mathcal{P}_{a} \ ,\ \mathcal{P}_{b}\right\} \ =\
\varkappa^2 M_{ab} \ + \ 2\, \Omega_{[ab]}{}^{c}{\mathcal{P}}_c\,,
\ee
\bea
\lb{PB-DD}
\left\{ \mathscr{D}_{\alpha} \ ,\ \mathscr{D}_{\beta}\right\} &=&{}
-4\varkappa M_{\alpha\beta} \ + \ 2i\, \Omega_{(\alpha\beta)}{}^{\gamma}{\mathscr{D}}_\gamma\,,  \\ [5pt]
\lb{PB-bDbD}
\left\{ \bar{\mathscr{D}}_{\dot\alpha} \ ,\ \bar{\mathscr{D}}_{\dot\beta} \right\} &=&{}
4\varkappa \bar M_{\dot\alpha\dot\beta} \ - \ 2i\, \Omega_{(\dot\alpha\dot\beta)\dot\gamma}\bar{\mathscr{D}}^{\dot\gamma}\,, \\ [5pt]
\lb{PB-DbD}
\left\{ \mathscr{D}_{\alpha} \ ,\ \bar{\mathscr{D}}_{\dot\beta} \right\} &=&{}
-2i(\sigma^c)_{\alpha\dot\beta}\mathcal{P}_{c} \ + \ i\, \Omega_{\dot\beta\alpha}{}^{\gamma}{\mathscr{D}}_\gamma
\ - \ i\, \Omega_{\alpha\dot\beta\dot\gamma}\bar{\mathscr{D}}^{\dot\gamma}\,,
\eea
\bea
\lb{PB-PD}
\left\{ \mathcal{P}_{a} \ ,\ \mathscr{D}_{\alpha}\right\} &=&{}
\frac{\varkappa}2\, (\sigma_a)_{\alpha\dot\gamma}\bar{\mathscr{D}}^{\dot\gamma} \ + \
\Omega_{a\alpha}{}^{\beta}{\mathscr{D}}_\beta \ - \ i\,
\Omega_{\alpha a}{}^{b}\mathcal{P}_b \,, \\ [5pt]
\lb{PB-PbD}
\left\{ \mathcal{P}_{a} \ ,\ \bar{\mathscr{D}}_{\dot\alpha} \right\} &=&{}
\frac{\varkappa}2\, (\sigma_a)_{\gamma\dot\alpha}{\mathscr{D}}^{\gamma} \ - \
\Omega_{a\dot\alpha\dot\beta}\bar{\mathscr{D}}^{\dot\beta} \ - \ i\,
\Omega_{\dot\alpha a}{}^{b}\mathcal{P}_b\,.
\eea

For the quantity \p{def-Pi} the Poisson brackets are
\be
\lb{PB-PiPi}
\left\{ \Pi_{a} \ ,\ \Pi_{b}\right\} \ =\
2\, \Omega_{[ab]}{}^{c}\Pi_c\,,
\ee
\bea
\lb{PB-PiD}
\left\{ \Pi_{a} \ ,\ \mathscr{D}_{\alpha}\right\} &=&{}
\frac{\varkappa}2\, (\sigma_a)_{\alpha\dot\gamma}\bar{\mathscr{D}}^{\dot\gamma} \ + \
\Omega_{a\alpha}{}^{\beta}{\mathscr{D}}_\beta \ - \ i\,
\Omega_{\alpha a}{}^{b}\Pi_b \,, \\ [5pt]
\lb{PB-PibD}
\left\{ \Pi_{a} \ ,\ \bar{\mathscr{D}}_{\dot\alpha} \right\} &=&{}
\frac{\varkappa}2\, (\sigma_a)_{\gamma\dot\alpha}{\mathscr{D}}^{\gamma} \ - \
\Omega_{a\dot\alpha\dot\beta}\bar{\mathscr{D}}^{\dot\beta} \ - \ i\,
\Omega_{\dot\alpha a}{}^{b}\Pi_b\,.
\eea

Also, one can calculate the following Poisson brackets:
\bea
&&
\left\{ \xi^{\beta} \ ,\ \mathscr{D}_{\alpha}\right\}\ = \ i\, \Omega_{\alpha\gamma}{}^{\beta}\xi^{\gamma}\,,
\qquad \left\{ \xi^{\beta} \ ,\ \bar{\mathscr{D}}_{\dot\alpha}\right\}\ = \ i\, \Omega_{\dot\alpha\gamma}{}^{\beta}\xi^{\gamma}\,,
\\ [5pt]
&&
\left\{ \bar\xi^{\dot\beta} \ ,\ \mathscr{D}_{\alpha}\right\}\ = \ i\, \Omega_{\alpha}{}^{\dot\beta}{}_{\dot\gamma}\bar\xi^{\dot\gamma}\,,
\qquad \left\{ \bar\xi^{\dot\beta} \ ,\ \bar{\mathscr{D}}_{\dot\alpha}\right\}\ = \ i\,
\Omega_{\dot\alpha}{}^{\dot\beta}{}_{\dot\gamma}\bar\xi^{\dot\gamma}\,,
\\ [5pt]
&&
\left\{ \xi^{\beta} \ ,\ \Pi_{a}\right\}\ = \ \Omega_{a\gamma}{}^{\beta}\xi^{\gamma}
- \frac{\varkappa}{2}\,(\bar\pi\tilde\sigma_a)^{\beta} \,,
\quad
\left\{ \bar\xi^{\dot\beta} \ ,\ \Pi_{a}\right\}\ = \ \Omega_{a}{}^{\dot\beta}{}_{\dot\gamma}\bar\xi^{\dot\gamma}
- \frac{\varkappa}{2}\,(\tilde\sigma_a \pi)^{\dot\beta}\,.
\eea

These relations are used in the analysis of the constraints in the model under consideration.

\end{document}